\documentclass{JHEP3} 

\usepackage{array}
\usepackage{float}
\usepackage{graphicx}
\usepackage{nicefrac}
\usepackage{amsmath,epsfig}
\usepackage{amssymb,amsfonts,mathrsfs}
\usepackage{color}
\usepackage{subfigure}
\usepackage{skak}
\let\normalcolor\relax
\usepackage{color}
\usepackage[T1]{fontenc}
\usepackage{mathtools} 
\usepackage{cite}

\usepackage{graphics}
\usepackage{color}
\usepackage{amsfonts}
\newcommand*{\bfrac}[2]{\genfrac{}{}{0pt}{}{#1}{#2}}

\newcommand{\newg}{{\bf{g}}}

\def\bea{\begin{eqnarray}}
\def\eea{\end{eqnarray}}
\def\be{\begin{equation}}
\def\ee{\end{equation}}
\def\ba{\begin{align}}
\def\ea{\end{align}}

\providecommand{\tabularnewline}{\\}

\def\tr{{\rm Tr}}

\title{\begin{center}
Integrability in 
$\mathcal{N}=2$ superconformal gauge theories
\end{center}}
\preprint{DESY 13 - 175}
\author{Elli Pomoni\footnote{Email: elli.pomoni@desy.de}
\\
\\
\it  DESY Theory Group, DESY Hamburg
\\
\it Notkestrasse 85, D-22603 Hamburg, Germany
\\
\\
\it Physics Division, National Technical University of Athens,
\\
\it
15780 Zografou Campus, Athens, Greece

\bigskip

\bigskip
}

\bigskip

\abstract{
\bigskip

Any ${\cal N}=2$ superconformal gauge theory (including
${\cal N}=4$ SYM) contains a set of local operators made only out of fields in the ${\cal N}=2$ vector multiplet  that is closed under renormalization to all loops, namely the $SU(2,1|2)$ sector.
For planar ${\cal N}=4$ SYM the spectrum of local operators can be obtained by mapping the problem to an integrable model (a spin chain in perturbation theory),
in principle for any value of the coupling constant.
We  present a diagrammatic argument that for any planar ${\cal N}=2$ superconformal gauge theory the $SU(2,1|2)$ Hamiltonian acting on infinite spin chains is identical  to all loops to that of ${\cal N}=4$ SYM, up to a redefinition of the coupling constant. Thus,  this sector is  integrable and anomalous dimensions can be, in principle, read off from the ${\cal N}=4$ ones up to this redefinition.

}

\begin{document}
\section{Introduction}

	After the discovery of the AdS/CFT correspondence, theoretical physics is experiencing a great upheaval. In particular, thanks to integrability, localization and dual string descriptions, we now possess a plethora of exact results in gauge theories that previously seemed unreachable \cite{Beisert:2010jr,Pestun:2007rz}. So far, the majority of these results has been only for the most symmetric gauge theory in four dimensions, namely the $\mathcal{N}=4$ SYM. It would be most unfortunate if these powerful techniques were to be valid only for this particular theory. We already know that localization  techniques are applicable in $\mathcal{N}=2$ supersymmetric gauge theories  \cite{Pestun:2007rz}, and we would like to investigate which other methods are transferable as well. In \cite{Grana:2001xn,Lin:2004nb,Gaiotto:2009gz,Gadde:2009dj,Colgain:2011hb,Aharony:2012tz,Stefanski:2013osa}, progress was made in figuring out the string dual to some $\mathcal{N}=2$ gauge theories.	
In the current article we want to investigate whether the property of integrability is present as well, building on work in \cite{Gadde:2010zi,Pomoni:2011jj,Gadde:2010ku,Liendo:2011xb,Gadde:2012rv,Korchemsky:2010kj,Belitsky:2004sf,Belitsky:2005bu,Beisert:2004fv,Ferretti:2004ba}.

\medskip

It would be very important to figure out which particular properties of a gauge theory make it integrable.
How necessary are {\it planarity, conformality, supersymmetry} (and how much supersymmetry) for the integrability of the $\mathcal{N}=4$ SYM theory?
Partial answers to these questions can be found in \cite{Zoubos:2010kh} and references therein as well as in \cite{Gadde:2010zi,Pomoni:2011jj,Gadde:2010ku,Liendo:2011xb,Poland:2011kg,Liendo:2011wc,Gadde:2012rv,Korchemsky:2010kj,Belitsky:2004sf,Belitsky:2005bu,Beisert:2004fv,Ferretti:2004ba}, where deformations of  $\mathcal{N}=4$ SYM  and other gauge theories with genuinely less supersymmetry are studied, respectively.  But a systematic approach is yet to be discovered.
Another critical point for the structure of the asymptotic Hamiltonian (dilatation operator) and thus for integrability is the representation of the fields under the color group.
As we saw clearly emerging  in the calculation of  \cite{Pomoni:2011jj},
 sectors with fields {\it only} in the vector multiplet enjoy Hamiltonians identical to the $\mathcal{N}=4$ SYM (up to two loops \cite{Pomoni:2011jj}, see also \cite{Korchemsky:2010kj,Belitsky:2004sf,Belitsky:2005bu,Beisert:2004fv,Ferretti:2004ba}), while sectors with bi-fundamental fields have Hamiltonians with deformed chiral functions\footnote{The simplest example of a chiral function is $\chi(1) = 1 -\mathbb{P}$.
As far as we know, the available calculations (see \cite{Sieg:2010jt} for a review) indicate that for $\mathcal{N}=4$ SYM only combinations of {\it undeformed chiral functions} appear in the Hamiltonian. However, when supersymmetry is lowered and bi-fundamantal fields are considered we encounter {\it deformed chiral functions} such as  $\chi(1) = 1 - \rho \,  \mathbb{P}$ \cite{Pomoni:2011jj}, with $\rho$ the ratio of the two coupling constants that correspond to the two color groups under which the  bi-fundamantal fields are charged.} \cite{Pomoni:2011jj}.

\medskip

 In order to face all these questions, we look at the next simplest cases (after $\mathcal{N}=4$), namely $\mathcal{N}=2$ superconformal gauge theories, and
 consider a particular, but quite large closed subsector  of the theory, specificaly the $SU(2,1|2)$ subsector which contains
 gauge-invariant local operators composed of
  fields only in the vector multiplet.
We argue that to {\it all-loop orders} in perturbation theory the {\it planar} and {\it asymptotic} Hamiltonian acting on states in this subsector is, up to a functional redefinition of the `t Hooft coupling constant\footnote{The `t Hooft coupling constant $\lambda = g^2_{YM}N = 16\pi^2 g^2$.} $g^2 \rightarrow f(g^2)= g^2 + \mathcal{O}\left(g^6\right)$, identical to the integrable dilatation operator of planar $\mathcal{N}=4$ SYM,
\be
\label{statement}
H_{\mathcal{N}=2}(g) = H_{\mathcal{N}=4}(\newg) \qquad \mbox{with} \qquad \newg = \sqrt{f(g^2)}   \, .
\ee 

The {\it planar} Hamiltonian (dilatation operator or mixing matrix) for a given sector (set of gauge-invariant local operators) at $\ell$ loops is obtained by computing the overall UV divergent piece of the two-point function
\be
\langle \bar{\mathcal{O}}_1(x) \mathcal{O}_2(y) \rangle^{(\ell)}
\ee
for all local operators $\mathcal{O}_1(x), \mathcal{O}_2(x)$ in the sector, to order $g^{2 \ell}$, in the large $N$ limit.\footnote{
In dimensional reduction \cite{Siegel:1979wq} it is extracted from the coefficient of the simple pole ($1/\varepsilon$). The finite piece of the two point functions affects only the normalization.
}
We refer the reader to the reviews
\cite{Minahan:2010js,Sieg:2010jt}
for more details.
In the spin chain picture each operator $\mathcal{O}(x)$ composed of $L$ fields corresponds to a spin chain state with $L$ sites. At each site the state space that we will consider is the infinite dimensional module $\mathcal{V} 
$
presented in section \ref{sector}.
The Hamiltonian can be thought of as a matrix acting on the total space $\oplus_{L=2}^{\infty} \mathcal{V}^{\otimes L}$.

\medskip

As an organizing principle we find it useful to think that we are first computing all
connected  off-shell $n$-point functions $G_{c \,n}^{(\ell)}$, then we insert them between the two {\it bare} operators  $\mathcal{O}(y)$ and $\bar{\mathcal{O}}(x)$ and finally Wick-contract the external legs of $G_{c \,n}$ with the {\it bare} operators. Schematically,\footnote{Note that apart from derivatives, the matrix structure of the Hamiltonian can already be read off from the connected $n$-point functions $G_{c \,n}$ that we insert in the two point function \eqref{gconin2pt}.}
\be
\label{gconin2pt}
\langle \bar{\mathcal{O}}(x) \mathcal{O}(y) \rangle^{(\ell)} \equiv \langle \bar{\mathcal{O}} (x)|G_{c \,n}^{(\ell)}(x,y) |  \mathcal{O}(y) \rangle \, .
\ee
Disconnected diagrams (after stripping off the composite operators) do not contribute to the Hamiltonian, because
 their overall divergencies only contain higher degree poles ($1/\varepsilon^n$ with $n>1$) \cite{Sieg:2010tz}.

\bigskip

The contents of the rest of paper are as follows.
We finish the introduction with an outline of the all-loop argument. We then begin the main bulk of the paper with section \ref{sector} and a description of the $SU(2,1|2)$ sector. In section
\ref{Language}
we introduce all the notation and the language that we will use throughout the paper.
Even though the statement \eqref{statement} holds for any $\mathcal{N}=2$ superconformal quiver, for simplicity we always think in terms of the $\mathcal{N}=2$ $SU(N)\times SU(N)$ elliptic quiver (the {\it interpolating theory}) that we describe in section \ref{interpolating}.
In section
\ref{difference} we review and elaborate on the diagrammatic observation that was made in \cite{Pomoni:2011jj} and led us to this paper.
In section \ref{123} we present and study
 explicitly the diagrams that lead to the one-, two- and three-loop Hamiltonians. Careful observation of these diagrams and comparison with the $\mathcal{N}=4$ SYM ones allows us to conclude that, to three loops, the statement \eqref{statement} is true.
Finally, in section \ref{All-loops} we use the lessons we learned by studying the diagrams up three loops to argue that \eqref{statement} should also hold as an  all-loop statement.
Some extra examples of multi-vertex insertions at four-, five- and six-loop are presented in Appendix \ref{multi-app} and two examples of powercounting  in Appendix \ref{non-ren-ex}.

\bigskip

\subsection{Outline of the argument}
\label{summary}

The main goal of this article is to argue that \eqref{statement} is a true statement for the all-loop Hamiltonian $H_{\mathcal{N}=2}(g)$ that acts in the  $SU(2,1|2)$  subsector (see \eqref{sector-fields}) of $\mathcal{N}=2$ superconformal gauge theories.
Our strategy is to study the difference
\be
\delta H^{(\ell)} = H_{\mathcal{N}=2}^{(\ell)} - H_{\mathcal{N}=4}^{(\ell)}  \, 
\ee
order by order in perturbation theory.   To obtain the Hamiltonian we need to compute all the connected  off-shell $n$-point functions of the $\mathcal{N}=2$ chiral superfield strength\footnote{The $\mathcal{N}=2$ chiral superfield strength $\mathcal{W}$ is the  $\mathcal{N}=2$ superfield that contains all the fields in the  $\mathcal{N}=2$ vector multiplet and thus all the fields in the  $SU(2,1|2)$  subsector \eqref{sector-fields} as its components \eqref{Wcomponents}. A collection of the basic superspace ingredients we use can be found in section \ref{superspace}.}
\be
\label{GcV}
\delta G_{c \,n}^{(\ell)}(\mathcal{W}) =  \langle \bar{\mathcal{W}}_1 \cdots {\mathcal{W}}_n\rangle^{(\ell)}_{\mathcal{N}=2}- \langle \bar{\mathcal{W}}_1 \cdots {\mathcal{W}}_n\rangle^{(\ell)}_{\mathcal{N}=4} \quad ,  \qquad \mathcal{W}_i\equiv \mathcal{W}(x_i,\theta_i,\tilde{\theta}_i)
\ee 
at loop order $\ell$, insert them in the two-point function $\langle \bar{\mathcal{O}}(x)\mathcal{O}(y)\rangle$ and Wick-contract, as we have schematically depicted in \eqref{gconin2pt}.
To simplify this complicated problem we
 organize our arguments in terms of the difference of the two effective actions 
\be
\label{Gamma2-4}
\delta \Gamma =  \Gamma_{\mathcal{N}=2} -  \Gamma_{\mathcal{N}=4}  \, ,
\ee
which is the generating functional of the difference of all the $n$-point 1PI functions that are relevant for the $SU(2,1|2)$ sector,
\be
\label{Gamma-n}
\delta \Gamma_n(\mathcal{W})  =  \Gamma^{\mathcal{N}=2}_n(\mathcal{W})  -  \Gamma^{\mathcal{N}=4}_n(\mathcal{W})  \, .
\ee
The purpose of the notation  $\Gamma_n(\mathcal{W})$ is to remind us that we {\it only} need 1PI diagrams whose external legs are ${\mathcal{W}}$'s and $\bar{\mathcal{W}}$'s and {\it not} bi-fundamental hypermultiplets.
This is a highly non-trivial statement which we discuss in section \ref{1PI2connected}.
There, we provide evidence that one can obtain all
 the connected off-shell $n$-point functions $\delta G_{c \,n}(\mathcal{W})$ that are relevant for the $SU(2,1|2)$ sector from  the  $n$-point 1PI functions $\delta \Gamma_n(\mathcal{W})$ \eqref{Gamma-n} alone.

\medskip

For our purposes it is useful to think of  the difference of the two effective actions \eqref{Gamma2-4} as the sum
\be
\delta \Gamma = \delta \Gamma_{ren. \, tree} + \delta  \Gamma_{new}
\ee
where $\delta  \Gamma_{ren. \, tree}$ denotes terms which were already present in the classical action and are now renormalized,  while $\delta  \Gamma_{new}$  {\it new effective vertices} which are created for the first time at some loop order. The $\delta  \Gamma_{ren. \, tree}$ terms yield the $n$-point 1PI functions that we will call {\it dressed skeletons} ({\it bare skeletons} that appeared at {\it tree level}\footnote{After stripping off the operators. See section \ref{skeletons} for a discussion on this language.} and are dressed at higher loops), while the $\delta  \Gamma_{new}$ terms lead to {\it new} $n$-point 1PI functions that were not there before.
The most crucial (non-trivial) step of our argument is that
the \textit{\textbf{new effective vertices $\delta  \Gamma_{new}$ cannot contribute to the renormalization of the operators of the $SU(2,1|2)$ sector}}, due to the following reasons:
\begin{itemize}
\item the {\it choice of the sector},
\item {\it planarity},
\item {\it Lorentz invariance}, 
\item a {\it non-renormalization theorem}  \cite{Fiamberti:2008sh,Sieg:2010tz}.
\end{itemize}
This crucial step of our argument is taken in sections \ref{classification} and
\ref{non-renorm-theorem}, while particular examples of $\delta  \Gamma_{new}$ vertices {\it not} contributing to the Hamiltonian are already encountered at two and tree loops in sections \ref{Two-loops} and \ref{Three-loops}, respectively.
The connected graphs $\delta G^{new}_{c \,n}(\mathcal{W})$ that are made out of $\delta  \Gamma_{new}$ can either not be planarly Wick-contracted to $\mathcal{O}\in SU(2,1|2)$ or (if they can) do not lead to logarithmic divergencies
\be
\label{finiteOper}
\langle \bar{\mathcal{O}} |\delta G^{new}_{c \,n}(\mathcal{W}) |  \mathcal{O} \rangle = finite  \qquad \forall \,  \mathcal{O} \in SU(2,1|2) \, ,
\ee
due to the {\it non-renormalization theorem} of \cite{Fiamberti:2008sh,Sieg:2010tz} that we discuss in section \ref{non-renorm-theorem}.
The first of the points above, ``{\it the choice of the sector}'' refers to the fact that gauge-invariant local operators $\mathcal{O} \in SU(2,1|2)$ are made {\it only}  out  of fields in the vector multiplet and have the spinor indices $\alpha$ always in the symmetric  representation of the $SU(2)_\alpha\in SU(2,2|2)$. ``{\it Lorentz invariance}''  refers to the fact that all the vertices in the effective action are Lorentz invariant (singlet/antisymmetric representation of the $SU(2)_\alpha$) while all the operators $\mathcal{O} \in SU(2,1|2)$ transform in the symmetric representation of the $SU(2)_\alpha$ and thus cannot be Wick-contracted. This is demonstrated with an explicit  example in equations \eqref{example1} and   \eqref{example2}.

\bigskip

As we discuss in section \ref{classification}, showing that
the {new effective vertices}  $\delta  \Gamma_{new}$ {cannot contribute to the renormalization of the operators of the $SU(2,1|2)$ sector}
 is feasible because for $\mathcal{N}=2$ superconformal theories there exists a {classification of all possible new terms} that can appear in the effective action.
 This classification of all possible new terms is obtained using {\it superconformal invariance} \cite{Buchbinder:1999jn} and
makes attainable the task of considering them all. 
Then, using the background field method (section \ref{BFF}), 
\begin{itemize}
\item {\it Gauge invariance}  (and {\it renormalizability})
\item  {\it $\mathcal{N}=2$ supersymmetry}
\end{itemize}
 force all the effects of the loops to be encoded in a single renormalization factor $Z(g)$. Combining all these facts,
we arrive, in section \ref{only-skeletons}, at the relation
\be
\label{tree-renorm-all}
\delta  \Gamma_{contributing}(\mathcal{W};g)    = \delta  \Gamma_{ren. \, tree}(\mathcal{W};g)   =  S_{tree} (\mathcal{W};\newg )\, .
\ee
This is a schematic equation that should be read as follows:  {\it the only 1PI diagrams that contribute to the Hamiltonian are dressed skeleton diagrams} with $\delta  \Gamma_{ren. \, tree}$ vertices! 
Although initially we make this statement for the 1PI's, in section \ref{1PI2connected} we will explain the reason why only these single vertices are enough to obtain all the connected diagrams that contribute to the Hamiltonian.
All this is done firstly for operators without derivatives,
for which the {matrix structure of the Hamiltonian} can already be seen in the connected $n$-point functions $G_{c \,n}$ \eqref{gconin2pt}.
Finally, in section \ref{derivatives-all-loop}  we observe that operators with derivatives in the $SU(2,1|2)$ sector 
create in the numerators of the loop integrals
{\it  traceless symmetric products of momenta}  that {\it do not alter their degree of divergence}. 
This concludes our argument for $\mathcal{N}=2$ superconformal gauge theories.

\bigskip

\section{The $SU(2,1|2)$  sector}
\label{sector}

In \cite{Pomoni:2011jj} we made a very simple but important observation: operators which mix in sectors including fields only in the vector multiplet, up to two-loops, enjoy Hamiltonians identical to the $\mathcal{N}=4$ one.  
This type of observation has also been made in  \cite{Korchemsky:2010kj,Belitsky:2004sf,Belitsky:2005bu} but was never really appreciated or put in use.
Before trying to use and generalize this observation we want to ask what is the biggest possible sector  made out  only of fields in the vector multiplet.

\medskip

The biggest sector of operators that are made  only out of fields in the ${\cal N}=2$ vector multiplet and that is closed to all loops is the  $SU(2,1|2)$  sector:
 \be
 \label{sector-fields}
\phi    \, ,  \quad    \lambda^{\mathcal{I}}_{+}      \, ,    \quad   \mathcal{F}_{++}  
  \, ,    \quad 
 \mathcal{D}_{+ \dot\alpha}  \, .
\ee
Above, for simplicity we choose  $\alpha = +$  in order to get the highest-weight state of the  {\it symmetric representation} of  the $SU(2)_\alpha$ part of the Lorentz group. But, of course, all the statements that we will make below hold for  {\it any} element in  the {symmetric 
representation}. Also, $\mathcal{I}=1,2$ is the $SU(2)_R$ symmetry index.

\medskip

The way to prove that the sector is closed to all loops goes as follows.
All the fields \eqref{sector-fields} that compose the operators $\mathcal{O}$ in this sector obey the condition 
\be
\label{sectorcong}
\Delta = 2 j - r    \qquad \forall \,  \mathcal{O} \in SU(2,1|2) \, ,
\ee
where $\Delta$ is the {classical} conformal dimension of the component fields, $r$ the {classical} $U(1)_r$ R-symmetry  and $j$ the {classical} $SU(2)_{\alpha}$ spin.
 In the conventions of \cite{Gadde:2009dj},  the $U(1)_r$ charges for the fields in \eqref{sector-fields} are $\left[ \phi \right]_r = -1$, $\left[ \lambda \right]_r = -1/2$ and $\left[ \mathcal{F} \right]_r = 0$.
Moreover, all the other individual fields (and  combinations of them) break this condition, and operators which contain them obey the inequality
\be
\label{outofsectorcong}
\Delta > 2 j - r    \qquad \forall \,  \mathcal{O} \notin SU(2,1|2) \, .
\ee
 For example, in the conventions of \cite{Gadde:2009dj} the $U(1)_r$ charge of  $\left[ \psi \right]_r = +1/2$ and its bare conformal dimension $\Delta_\psi = 3/2 > 2 j - r = 1/2$.
 The condition \eqref{sectorcong} is not a BPS condition. $\Delta$, $r$ and $j$ will be corrected in perturbation theory, but they will never be corrected enough to mix with the rest of the fields that break \eqref{sectorcong} by an integer.
Thus,  in perturbation theory the $SU(2,1|2)$ sector is closed to all orders in the `t Hooft coupling constant, and since the `t Hooft coupling
expansion is believed to converge \cite{'tHooft:1982tz}, this statement is also true for any finite value of the `t Hooft coupling constant in the planar limit.

\medskip

 It is also very simple to check that all the fields \eqref{sector-fields} and thus all the operators in the $SU(2,1|2)$ sector obey
 \be
\Delta = 2 +r - 2\bar{j}
\qquad
\mbox{and}
\qquad
 j+\bar{j} = 1 +r    \qquad \forall \,  \mathcal{O} \in SU(2,1|2) \, .
 \ee
The $SU(2,1|2)$ sector includes many smaller familiar subsectors such as $SU(1|1)$ and $SU(1,1)$.

\medskip

Finally, we would also like to note that, if the incoming operator is $\mathcal{O} \in SU(2,1|2)$, the outgoing (conjugate) operator 
 \be
  \bar{\mathcal{O}} \in SU(1,2|2) \qquad  \mbox{is made out of} \qquad \bar{\phi}   \, ,    \quad     \bar{\lambda}_{\dot+ \,\mathcal{I}}   \, ,    \quad   \bar{\mathcal{F}}_{\dot+\dot+}    \, ,    \quad 
 \mathcal{D}_{\alpha \dot+}
\, \, .
 \ee
The outgoing operator $\bar{\mathcal{O}}$ is in the {\it symmetric  representation} of the $SU(2)_{\dot\alpha}$ part of the Lorentz group and it obeys
\be
\label{sectorcongbar}
\Delta  = 2 \bar{j} + r    \qquad \forall \,  \bar{\mathcal{O}} \in SU(1,2|2) \, .
\ee

\smallskip

%
%
%
%

In the spin chain picture each operator $\mathcal{O}(x)$ composed of $L$ fields corresponds to a spin chain state with $L$ sites. At each site, the state space that corresponds to the $SU(2,1|2)$ subsector is the infinite dimensional module
 $\mathcal{V} = \{\mathcal{D}^n_{+
\dot{\alpha}} \left( \phi \, , \, \lambda_{+}^{\mathcal{I}} \, , \, \mathcal{F}_{++} \right)\}
$
 where $n=0,\dots,\infty$ is the number of derivatives at each site.
  The $SU(2,1|2)$ sector is a non-compact analogue of the $SU(3|2)$ sector of Beisert \cite{Beisert:2003ys}. Let us see how this arises. 
We begin the study of spin chains for $\mathcal{N}=2$ superconformal gauge theories by identifying the equivalent of the BMN vacuum. For every color group in the quiver we can consider a ``holomorphic'' operator $\tr \phi^{L}$, made out of the complex scalar in the vector multiplet, that is part of the chiral ring and thus protected (has zero anomalous dimension). This operator will always obey the BPS condition $\Delta = r$ (=$L$). We define $\Delta - r$ as the magnon number. The operator $\tr \phi^{L}$ has $\Delta - r=0$, so the spin chain state to which it corresponds can be considered as the vacuum. 
The next step is to identify the single magnon states with $\Delta - r=1$.
 One can go through all the fields available in $\mathcal{N}=2$ superconformal gauge theories and discover that the only operators with $\Delta - r=1$ are the ones with 
  a single insertion in the sea of $\phi$'s of a $\left( \lambda_\alpha^{\mathcal{I}}, \mathcal{D}_{\alpha \dot\alpha}\right)$  with indices in the adjoint of the color group or a bi-fundamental $(Q^{\mathcal{I}},\bar{\psi}_{\dot\alpha})$\footnote{When the bi-fundamentals $(Q^{\mathcal{I}},\bar{\psi}_{\dot\alpha})$'s are inserted they interpolate between two different vacua and the spin chain for a single magnon can never be closed \cite{Gadde:2010zi}.}. All the other states (operators) with $\Delta - r>1$ should be, from the spin chain point of view, thought of as composite states (bound states) made out of the elementary (single magnon) excitations $\left( \lambda_\alpha^{\mathcal{I}}, \mathcal{D}_{\alpha \dot\alpha}\right)$  and $(Q^{\mathcal{I}},\bar{\psi}_{\dot\alpha})$.

  \bigskip
  
Following Beisert \cite{Beisert:2005tm},   the choice of the BMN vacuum $\tr \phi^{L}$ breaks the symmetry as follows
\be
SU(2,2|2) \rightarrow SU(2|2)_R \times SU(2)_{\alpha}  \nonumber \, .
\ee
In other words $SU(2|2)_R \times SU(2)_{\alpha}$ is the symmetry that the single magnon excitations $\left( \lambda_\alpha^{\mathcal{I}}, \mathcal{D}_{\alpha \dot\alpha}\right)$  and $(Q^{\mathcal{I}},\bar{\psi}_{\dot\alpha})$ enjoy in the $\phi$-vacuum. In our notations $SU(2|2)_R$ has $SU(2)_R$ and $SU(2)_{\dot\alpha}$ as its bosonic subgroups.
Excitations that correspond to  broken generators $\left( \lambda_\alpha^{\mathcal{I}}, \mathcal{D}_{\alpha \dot
\alpha}\right)$  will become {\it Goldstone excitations}  and they will be the {\it gapless magnons}\footnote{The magnon that do not come from broken generators are gapped magnons $(Q^{\mathcal{I}},\bar{\psi}_{\dot\alpha})$ (non-Goldstone excitations) \cite{Gadde:2010ku}.} according to Table \ref{spinchainsymmetry}.
\begin{table}[h!!]
\begin{centering}
\begin{tabular}{ccc|c}
 & $SU(2)_{\dot{\alpha}}$ & \multicolumn{1}{c}{$SU(2)_{R}$} & $SU(2)_{\alpha}$ \tabularnewline
\cline{2-3} 
\multicolumn{1}{c|}{$SU(2)_{\dot{\alpha}}$} & \multicolumn{1}{c|}{$\mathcal{L}_{\:\dot{\beta}}^{\dot{\alpha}}$} & $\mathcal{Q}_{\: \mathcal{J}}^{\dot{\alpha}}$ & $\mathcal{\mathcal{D}}_{\:\beta}^{\dagger\dot{\alpha}}$ \tabularnewline
\cline{2-3} 
\multicolumn{1}{c|}{$SU(2)_R$} & \multicolumn{1}{c|}{$\mathcal{S}_{\:\dot{\beta}}^{\mathcal{I}}$} & $\mathcal{R}_{\: \mathcal{J}}^{\mathcal{I}}$ & $\lambda_{\:\beta}^{\dagger \mathcal{I}}$
\tabularnewline
\cline{2-4} 
$SU(2)_{\alpha}$ & $\mathcal{D}_{\:\dot{\beta}}^{\alpha}$ & $\lambda_{\: \mathcal{J}}^{\alpha}$ & \multicolumn{1}{c|}{$\mathcal{L}_{\:\beta}^{\alpha}$} 
\tabularnewline
\cline{4-4} 
\end{tabular}
\par\end{centering}
\caption{\it The symmetry structure of the $\mathcal{N}=2$ quiver spin chains. The choice of the vacuum  breaks the $SU(2,2|2)$ symmetry down to
$SU(2|2)_R \times SU(2)_{\alpha}$. The  $SU(2)_R$ and $SU(2)_{\dot\alpha}$ are the bosonic subgroups of $SU(2|2)_R$. The broken generators become the {\it Goldstone excitations} $\left( \lambda_\alpha^{\mathcal{I}}, \mathcal{D}_{\alpha \dot
\alpha}\right)$.
}
\label{spinchainsymmetry}
\end{table}
\\
The Goldstone excitations $\left( \lambda^{\mathcal{I}}_\alpha, \mathcal{D}_{\alpha \dot
\alpha}\right)$ belong to the same supersymmetry multiplet
\be
\mathcal{Q}_{\dot\beta}^{\mathcal{I}}  \, |\lambda_{\alpha}^{\mathcal{J}}\rangle  = \epsilon^{\mathcal{I}\mathcal{J}} \,  |\mathcal{D_{\alpha\, \dot\beta}}\rangle \, .
\ee
As such, they must have the same dispersion law
\be
E(p;g)=  \sqrt{1+8 f(g^2)\sin^{2}\left(\frac{p}{2}\right)  }
\ee
that is fixed by the $SU(2|2)_R$ symmetry up to an unknown function $f(g^2)$ of the coupling constants \cite{Beisert:2005tm},
\be
f(g^2) = g^2 + \mathcal{O}(g^4) \, .
\ee
After fixing the spin $SU(2)_\alpha$ quantum number to $\alpha = +$, its  highest weight state, the $SU(2|2)_R$ symmetry of the excitations around the vacuum $\tr \phi^{L}$ also fixes the scattering matrix  up to an unknown function $f(g^2)$ of the coupling constant \cite{Beisert:2005tm}.  The $SU(2|2)$ scattering matrix for the $\left( \lambda_+^{\mathcal{I}}, \mathcal{D}_{+ \dot\alpha}\right)$
immediately satisfies the Yang-Baxter equation (YBE) (given the fact that the $\mathcal{N}=4$ one does \cite{Beisert:2005tm}) and thus we expect the $SU(2,1|2)$ sector to be integrable.

\bigskip

We wish to conclude this section by presenting one more argument for the
integrability of the $SU(2,1|2)$ sector that is based on AdS/CFT ideology and was presented in \cite{Gadde:2012rv}.
For this argument we will consider a quite wide and well studied class of ${\cal N}=2$ superconformal theories that admit a string dual description: orbifolds of ${\cal N}=4$ SYM. When orbifolding ${\cal N}=4$ SYM by a discrete subgroup {$\Gamma \subset SU(2) \subset SU(4)_R$}, 
elliptic quivers with a product gauge group $SU(N)^M$ are obtained, with their $M$ gauge couplings being
exactly marginal parameters.  
The gravity duals of these ${\cal N}=2$ superconformal theories are described by type IIB string theory on  $AdS_5 \times S^5/\Gamma$   \cite{Kachru:1998ys, Lawrence:1998ja}.
  At  strong coupling one can compute the $SU(2|2)$
  S-matrix of the excitations around the BMN vacuum using the sigma model description. String states in the $SU(2,1|2)$ sector are classically described by the same sigma model as the ${\cal N}=4$ ones  because they live only in the $AdS_5 \times S^1$ part of the space, {\it i.e.} in directions of the target space unaffected by the orbifold.  To be more concrete, we consider the $\mathbb{Z}_2$  orbifold case which has two marginal 't Hooft couplings $\lambda$ and $\check \lambda$. 
  The dictionary between the gauge theory and the string theory parameters in this case is
 \begin{eqnarray} \label{Bdictionary}
&& \frac{1}{g_{YM}^2} +   \frac{1}{\check g_{YM}^2}   =  \frac{1}{2 \pi g_s} \, , \qquad \frac{\check g_{YM}^2}{g_{YM}^2}  = \frac{\beta}{1-\beta} \, , \qquad  \beta \equiv \int_{S^2} B_{NS}\,
 \end{eqnarray}
 where  $B$ is the the NSNS field with  period $\beta$ through the  blown-down $S^2$ of the orbifold singularity \cite{Lawrence:1998ja,Klebanov:1999rd}. 
  The only difference with  ${\cal N}=4$ SYM is that
 the relation between $\alpha'$ and the AdS radius $R$ changes to
 \be \label{tensionrenorm}
 f(g^2) = \frac{R^4}{\left( 2 \pi \alpha' \right)^2} = \frac{2 \lambda \check \lambda}{ \lambda + \check \lambda} \, .
 \ee
 Thus, the only difference of this $\mathcal{N}=2$  S-matrix from the $\mathcal{N}=4$ one will be a renormalization (rescaling) of the string tension. 
 Tempted by the fact that we can see integrability at one-and two-loops \cite{Pomoni:2011jj}, as well as at the strong coupling \cite{Gadde:2012rv}, we were motivated to look for integrability for any value of the couplings $\lambda$ and $\check \lambda$. 

\bigskip

\section{Language}
\label{Language}

As we will see throughout this article, it is very important to use the appropriate (convenient) language (formalism); the one that keeps manifest as much symmetry as possible.
When we use $\mathcal{N}=1$ superspace, $\mathcal{N}=1$ supersymmetry is manifest and to obtain $\mathcal{N}=2$ we need to further impose $SU(2)_R$. On the other hand, when we use $\mathcal{N}=2$ superspace the full $\mathcal{N}=2$ supersymmetry is  immediately  manifest.
For $\mathcal{N}=2$ theories it is preferable to perform calculations in  $\mathcal{N}=2$ superspace where even the intermediate steps of the calculation reflect the fact that $\mathcal{N}=2$ supersymmetry is  present. However, given the fact that the bigger part of the integrability community  is not used  to  $\mathcal{N}=2$ superspace, and also that we eventually want to generalize our argument to $\mathcal{N}=1$ theories\footnote{This is work in progress.}, we will initially make as many steps as possible using  $\mathcal{N}=1$ superspace.
What is more, background field formalism makes gauge invariance manifest and when combined with supersymmetry leads to very powerful non-renormalization theorems that explain many ``{\it miraculous cancellations}''\footnote{The phrase ``{\it miraculous cancellations}'' we borrow from the title
  ``Miraculous Ultraviolet Cancellations in Supersymmetry Made Manifest'' of the seminal work \cite{Howe:1983sr}.}  \cite{Grisaru:1979wc,Grisaru:1982zh,Grisaru:1980nk,Stelle:1981gi,Howe:1982tm,Howe:1983wj,Howe:1983sr}. 
For a more modern approach on the background field method (BFM) in $\mathcal{N}=2$ superspace the interested reader can see \cite{Buchbinder:1997ya,Buchbinder:1998np} for Harmonic and \cite{Jain:2013hua} for Projective superspace.

\bigskip
\subsection{Skeletons}
\label{skeletons}

\begin{figure}[h!!]
   \centering
      \includegraphics[scale=0.45]{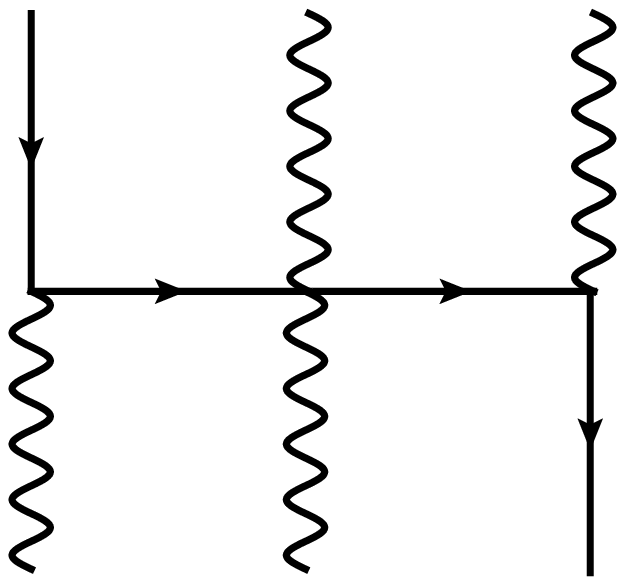}
          \put(-45.5,-15){{\footnotesize ($i$)}}
   \qquad    \qquad    \qquad
   \includegraphics[scale=0.45]{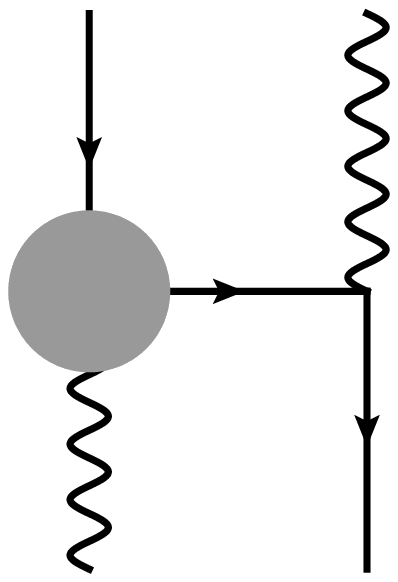}
          \put(-49,35){{\tiny 1-$loop$}}
          \put(-28.5,-15){{\footnotesize ($ii$)}}
   \qquad    \qquad    \qquad
   \includegraphics[scale=0.45]{self-energy.eps}
       \put(-20.5,35){{\tiny 2-$loop$}}
          \put(-19.5,-15){{\footnotesize ($iii$)}}
          \qquad    \qquad    \qquad
   \includegraphics[scale=0.45]{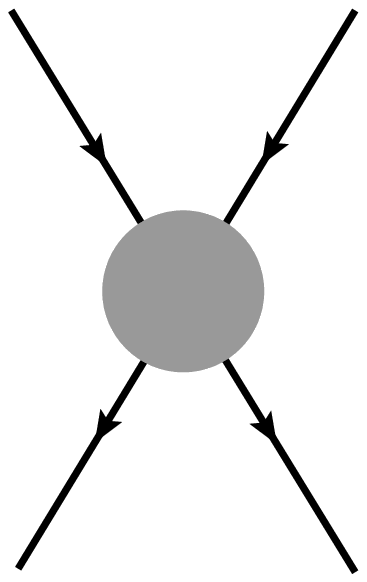}
          \put(-30,35){{\tiny $new$}}
                    \put(-28,-15){{\footnotesize ($iv$)}}
 \caption{\it At two-loops,  the  types of diagrams that contribute to the renormalization of operators are bare skeleton diagrams of order $g^4$,  one-loop self-energy and vertex  corrections to the tree-level skeleton diagrams of order $g^2$ that appeared in the one-loop order (dressed skeletons) and  two-loop self-energy corrections. What is more, operators can be renormalized by new vertices that appear for the first time in the effective action at one-loop.}
 \label{2-loop}
\end{figure}

All tree-level connected graphs we either call  {\it  tree-level  skeletons} or {\it bare  skeletons}. In Figures \ref{1-loop},  \ref{2-loop} and  \ref{3-loop} all the diagrams on the left hand side are   {\it bare skeleton diagrams}.
At higher loops, the  tree-level connected graphs ({\it  bare skeletons}) will be corrected ({\it dressed})   by connected propagators (self-energy) and vertex corrections. We will refer to these graphs  as 
 {\it dressed skeleton diagrams}. Examples of   {\it dressed skeletons} are depicted in  Figures \ref{1-loop}($i$),  \ref{2-loop}($ii$),  \ref{3-loop}($ii$) and \ref{3-loop}($iii$)  with the grey bubbles  denoting the vertex and leg corrections.

\medskip
 
 The graphs that contain {\it new vertices} $\Gamma^{new}_{n}$ in the effective action   -- with the  grey bubbles saying {\it new} --  will {\it not} be called skeletons!  See Figures \ref{2-loop} ($iv$),  \ref{3-loop} ($v$) and  \ref{3-loop} ($vi$).  The monicker ``skeletons'' refers only to diagrams that appeared already at tree level, while the graphs that correspond to $\Gamma^{new}_{n}$ are made out of loop diagrams.

\subsection{Superspace}
\label{superspace}
In this section we review some basic elements of the superspace formalism that we will use throughout the paper.  For a review see \cite{Gates:1983nr,Buchbinder:1995uq,Galperin:2001uw}.

\medskip

In $\mathcal{N}=1$ superspace language, the fields that compose the operators  in the  $SU(2,1|2)$  sector are components of the $\mathcal{N}=1$ vector superfield $V$ and of the $\mathcal{N}=1$ adjoint chiral superfield $\Phi$ that form the  $\mathcal{N}=2$ vector multiplet.
In the  Wess-Zumino gauge,
 \be
 \Phi|_{\theta =0} = \phi   \, ,  \quad D_+ \Phi |_{\theta=0} =  \lambda^{1}_{+}    \, ,  \quad W_+ |_{\theta=0} =  \lambda^{2}_{+}     \, ,    \quad    D_+W_+ |_{\theta=0}=  \mathcal{F}_{++} 
 \ee
 where $W_\alpha$ is the $\mathcal{N}=1$ chiral superspace field strength  $W_{\alpha} \equiv i  \bar{D}^2\left(e^{-V}D_\alpha e^{V} \right)$. The usual spacetime derivatives can be written as
 \be
i \, \mathcal{\partial}_{+ \dot\alpha}=   \left\{ D_+\, ,\, \bar D_{\dot\alpha}\right\} \, .
 \ee
 In order to obtain covariant derivatives, one goes through the definition of a gauge-covariant, super-covariant derivative 
 \be
 \mathcal{D}_\alpha = e^{-V} {D}_\alpha e^{V} \,  .
 \ee
 Note that in the $\mathcal{N}=1$ superspace language we need two different superfields to describe the $SU(2,1|2)$ sector. This increases the number of diagrams that have to be considered and obscures the computation for other reasons such as gauge fixing.
 
 \medskip

  In the most naive form of  $\mathcal{N}=2$  superspace language (namely the real superspace $\mathbb{R}^{4|8}$ with coordinates $\{x,\theta,\tilde\theta  \}$) the  $\mathcal{N}=2$ vector multiplet can be written using the  $\mathcal{N}=2$  chiral superfield strength \cite{Grimm:1977xp}
\be
\label{WN2}
\mathcal{W} = \Phi + \tilde\theta^\alpha W_\alpha +  \tilde\theta^2 G \, .
\ee
 All the fields in \eqref{sector-fields} can be obtained from the $\mathcal{N}=2$ field strength 
 \be
 \label{Wcomponents}
 \mathcal{W}|_{\theta =  \tilde{\theta}=0} = \phi   \, ,  \quad D_+^\mathcal{I} \mathcal{W} |_{\theta=  \tilde{\theta}=0} =  \lambda^{\mathcal{I}}_{+}      \, ,    \quad  D_+^\mathcal{I}   D_+\,_\mathcal{I} \mathcal{W} |_{\theta=  \tilde{\theta}=0}=  \mathcal{F}_{++}  
\ee
 in  Wess-Zumino gauge, and
\be
i\,  \delta^\mathcal{I}_{\mathcal{J}} \, \mathcal{ \partial}_{+ \dot\alpha}=   \left\{ D_+^\mathcal{I} \, ,\, \bar D_{\dot\alpha\, \mathcal{J}}\right\}  \, ,
 \ee
where $\mathcal{I}=1,2$ is the $SU(2)_R$ symmetry index.
Within this formalism, the $\mathcal{N}=2$ SYM classical Lagrangian can be compactly written as
\be
\label{classical-prepotential}
\mathcal{L}(\mathcal{W};g) = \frac{1}{g^2} \int  d^2\theta  d^2\tilde{\theta} \, \tr \big(\mathcal{W}^2\big) = 
 \frac{1}{g^2}  \left[ \int d^2\theta  \, \tr \big(W^{\alpha}W_{\alpha}\big)
+  \int  d^2\theta d^2\bar{\theta}  \, \tr \big( e^{-V} \bar{\Phi}e^{V}\Phi \big)   \right]\,.
\ee
Before concluding this section, we should stress that {the calculations should not be done in the Wess Zumino gauge}. The WZ gauge breaks supersymmetry and obscures the intermediate steps of the calculations \cite{Kovacs:1999rd}. What is more, going from $\mathcal{N}=2$ superspace to $\mathcal{N}=1$ we partially gauge fix. Gauge theory in $\mathcal{N}=2$ superspace enjoys a bigger gauge invariance than the $\mathcal{N}=1$ one.

\bigskip

\subsection{The interpolating theory}
 \label{interpolating}

 Before drawing Feynman diagrams we wish to explicitly give the Lagrangian of an example of an  $\mathcal{N}=2$ superconformal gauge theory, so that the reader can always have in mind what are the possible vertices that could be used and what are the Feynman rules. We pick to present the $\mathcal{N}=2$  $SU(N)\times SU({N})$  elliptic quiver which has two exactly marginal coupling constants $g$ and $\check{g}$ and is the conformal theory, considered in \cite{Gadde:2009dj,Gadde:2010zi,Pomoni:2011jj,Liendo:2011xb}, which interpolates between the $\mathcal{N}=2$ superconformal QCD (SCQCD)  (for $\check{g}=0$) and  the $\mathbb{Z}_2$ orbifold of  $\mathcal{N}=4$ that has the same dilatation operator with $\mathcal{N}=4$ SYM (for $\check{g}=g$). We like to refer to it as {\it the interpolating theory}.

\medskip

We will use the $\mathcal{N}=2$ elliptic quiver with two color groups $SU(N)\times SU(N)$  as the paradigmatic example, but our results are easily generalizable to any $\mathcal{N}=2$  superconformal quiver.
In particular we will argue that the statement \eqref{statement} is true for any $\mathcal{N}=2$ superconformal ADE quiver (that correspond to a finite or affine Dynkin diagram) with the nodes denoting the color groups and the lines that connect them the fundamentals or bifundamental matter (quarks).

\medskip

In terms of $\mathcal{N}=1$ superfields, 
with the conventions of \cite{Pomoni:2011jj}, the action reads
\begin{equation}\label{actionN2}
\begin{aligned}
S
&=\frac{1}{2}\int d^4xd^2\theta\Big[\frac{1}{g_{YM}^2}\tr\big(W^{\alpha}W_{\alpha}\big)
+\frac{1}{\check g_{YM}^2}\tr\big(\check W^{\alpha}\check W_{\alpha}\big)\Big]
\\
&+\int d^4x d^4\theta\big[\tr\big(e^{-g_{YM} V} \bar{\Phi}e^{g_{YM} V}\Phi\big)
+\tr\big(e^{-\check{g}_{YM} \check{V}} \bar{\check{\Phi}}e^{\check{g}_{YM} \check{V}}\check{\Phi}\big)\big]\\
&\phantom{{}={}}
+
\int d^4x d^4\theta\big[\tr\big(\bar{Q}^{\hat{\mathcal{I}}}e^{g_{YM} V}Q_{\hat{\mathcal{I}}}e^{-\check{g}_{YM} \check{V}}\big)
+\tr\big(\bar{\tilde{Q}}_{\hat{\mathcal{I}}}e^{\check g_{YM} \check V} \tilde{Q}^{\hat{\mathcal{I}}}e^{-g_{YM} V}\big)\big]\\
 &\phantom{{}={}}
+i\int d^4xd^2\theta\big[g_{YM}\tr\big(\tilde{Q}^{\hat{\mathcal{I}}} \Phi Q_{\hat{\mathcal{I}}}\big)-\check g_{YM}\tr\big(Q_{\hat{\mathcal{I}}} \check{\Phi}\tilde{Q}^{\hat{\mathcal{I}}}\big)\big]\\
&\phantom{{}={}}
-i\int d^4x d^2\bar\theta\big[g_{YM}\tr\big(\bar Q^{\hat{\mathcal{I}}}\bar\Phi\bar{\tilde{Q}}_{\hat{\mathcal{I}}}\big)-\check g_{YM}\tr\big(\bar{\tilde{Q}}_{\hat{\mathcal{I}}}\bar{\check{\Phi}}\bar Q^{\hat{\mathcal{I}}}\big)\big]   \, ,
\end{aligned}
\end{equation}
where now $W_{\alpha} \equiv i  \bar{D}^2\left(e^{-g_{YM}V}D_\alpha e^{g_{YM}V} \right)$.
The global 
$SU(2)_{\text{R}}$ symmetry that transforms the chiral quark $Q$ into the anti-chiral $\bar{\tilde{Q}}$
is not manifest in the $\mathcal{N}=1$ superspace language.
The field content  of the theory and its transformation properties 
under the $SU(N)\times SU({N})$ gauge and the 
$SU(2)_\text{R}\times U(1)$ R-symmetry groups
are shown in table \ref{tab:fcont}.
\begin{table}[h]
\begin{center}
\begin{tabular}{c|c|c|c}
$\text{field}$ & $SU(N)\times SU(\check{N})$  & $SU(2)_\text{R}$ & $U(1)$  \\
\hline
$V$ & $(\text{adj}.,1)$ & $1$ & $0$ \\
$\Phi$ & $(\text{adj}.,1)$ & $1$ &  $1$ \\
\hline
$\check{V}$ & $(1,\text{adj}.)$ & $1$ &  $0$\\
$\check{\Phi}$ & $(1,\text{adj}.)$ & $1$ &  $1$\\
\hline
$Q_{\hat{\mathcal{I}}}$ & $(\Box,\bar{\Box})$ &  $\Box$ & $0$ \\
$\tilde Q^{\hat{\mathcal{I}}}$ & $(\bar{\Box},\Box)$ &    $\bar{\Box}$& $0$ \\
\end{tabular}
\caption{The field content of the $\mathcal{N}=2$ \emph{interpolating theory} in terms of $\mathcal{N}=1$ superfields. \label{tab:fcont}}
\end{center}
\end{table}
The index $\hat{\mathcal{I}}=1,2$ denotes an extra $SU(2)_L$ global symmetry that {\it the interpolating theory} has for all values of the coupling constants \cite{Gadde:2009dj}. The relation between the Yang-Mills coupling constants, $g_{YM}$ and  $\check{g}_{YM}$,  that appear in the Lagrangian and the `t Hooft coupling constants that appear in the loop expansion of the dilatation operator is
\be
\lambda = g^2_{YM}N = 16\pi^2 g^2  \quad , \qquad \check{\lambda} = \check{g}^2_{YM}N = 16\pi^2 \check{g}^2  \, .
\ee
The explicit Feynman rules can be found in the Appendix of \cite{Pomoni:2011jj}.

\bigskip

\subsection{Background field formalism}
\label{BFF}

The gauge theories that we are considering are renormalizable theories and all the divergencies encountered will be renormalized, with all the loop effects encoded in $Z$'s that relate bare and renormalized quantities for every field and vertex that appear in the tree-level Lagrangian
\be
g_0 = Z_g \, g \, , \quad  V_0 = \sqrt{Z_V} \, V \, ,  \quad  \Phi_0 = \sqrt{Z_\Phi} \,  \Phi  \, ,  \quad  \left(W_\alpha \right)_0 = \sqrt{Z_{W_\alpha}} \,  W_\alpha  \, , \quad \alpha_0 = Z_\alpha\, \alpha \, ,
\ee
as well as for the gauge fixing parameter $\alpha$.
In principle we have to calculate all possible leg and vertex corrections (all possible  $G_{c \,n}$).
Connected $n$-point functions $G_{c \,n}$ with conventional gauge fixing do not obey  Ward identities (WI), but the more complicated  Slavnov-Taylor identities.
 The background field formalism allows gauge fixing without loosing explicit gauge invariance (see \cite{Abbott:1981ke} for a comprehensive review and \cite{Howe:1982tm,Howe:1983sr} for the superspace discussion). Connected $n$-point functions $G_{c \,n}$ of the background fields obey WI.
In $\mathcal{N}=1$ superspace language all we have is $Z_g(g) $, $Z_V(g) $ and  $Z_\Phi(g)$\footnote{Plus $Z_\alpha$ for the gauge fixing parameter, when we consider non gauge-invariant quantities.}. The action is a functional of the classical background fields for which
\be
\label{ZVG}
Z_g(g)  \sqrt{Z_{V}(g)} =1
\ee due to conservation of charge (the usual WI that is due to gauge invariance). There is a second Ward identity due to  $\mathcal{N}=2$  supersymmetry
\be
\label{ZWphi}
Z_{W_\alpha} (g) =  Z_\Phi(g)  = Z_{\mathcal{W}} (g)  \, ,
\ee
where $Z_{\mathcal{W}} (g)$ is the $Z$ factor of the $\mathcal{N}=2$ chiral superfield strength 
\be
\mathcal{W}_0 = \sqrt{Z_{\mathcal{W}}} \,  \mathcal{W}  \, ,
\ee 
defined in \eqref{WN2}.
However, this WI  only holds when the full $\mathcal{N}=2$ gauge symmetry is properly preserved.
Gauge fixing  $\mathcal{N}=2$ gauge invariance down to $\mathcal{N}=1$ will break  \eqref{ZWphi} if done in a crude way. This is why it is important to perform the calculations in $\mathcal{N}=2$ superspace where  \eqref{ZWphi} is automatic  \cite{Howe:1983sr}. The use of BFM guarantees \eqref{ZVG} and relates $Z_{\mathcal{W}}(g)$ to $Z_{g}(g)$, and
all in all there is {\it one function of the coupling $Z(g)$} that encodes all the information about divergencies and renormalization.

\medskip

Renormalization of composite operators, already a complicated problem itself, is further obscured
when done with conventional gauge fixing where 1PI $n$-point functions $\Gamma_n$  do not obey WI.
As a result of conventional gauge fixing, gauge-invariant operators mix with non-gauge-invariant operators of the same (classical) conformal dimension. All this complexity can be overcome  \cite{KlubergStern:1975hc} by using the BFM, that allows gauge fixing without loosing explicit gauge invariance.
For a rather modern and clear presentation of renormalization of composite operators in the background field gauge see \cite{Tarrach:1981bi}.

\bigskip

\section{The diagrammatic difference with $\mathcal{N}=4$ SYM}
\label{difference}

\begin{figure}[h!]
\begin{centering}
\includegraphics[scale=0.5]{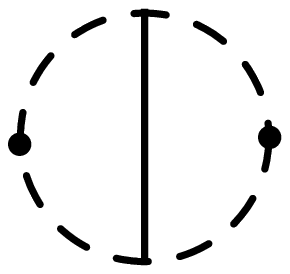}
\qquad   \qquad
\includegraphics[scale=0.5]{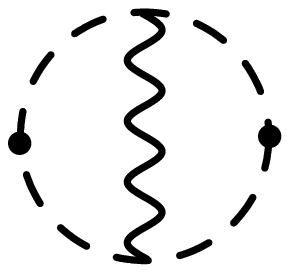}
\qquad
\includegraphics[scale=0.5]{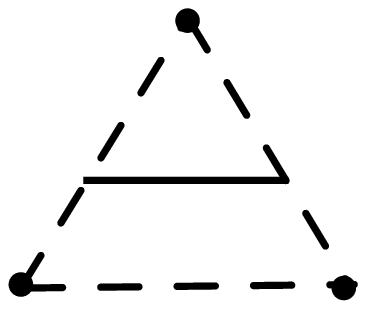}
\qquad
\includegraphics[scale=0.5]{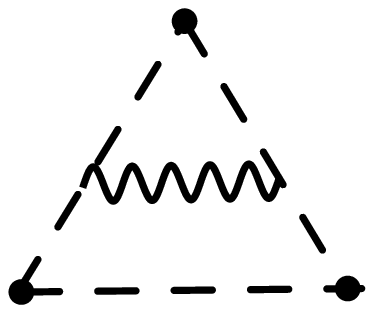}
\\
\, ~ \,
\\
\includegraphics[scale=0.5]{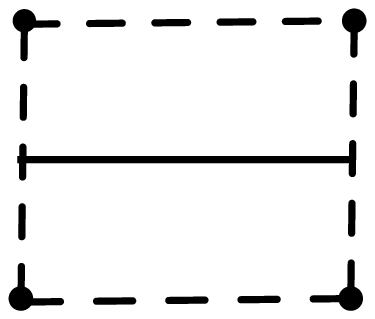}
\qquad
\includegraphics[scale=0.5]{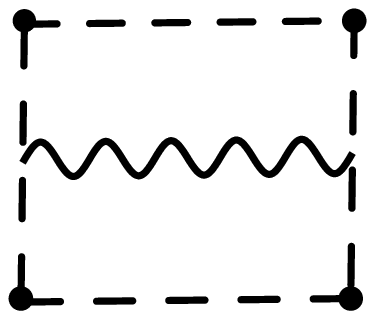}
\qquad
\includegraphics[scale=0.5]{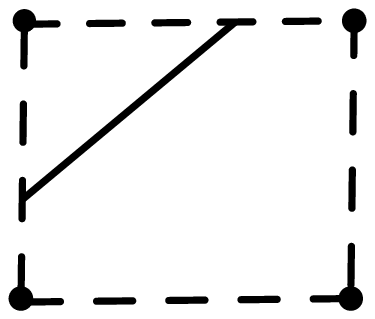}
\qquad
\includegraphics[scale=0.5]{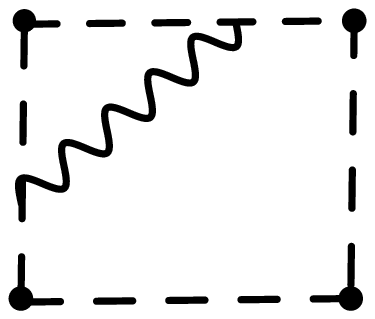}
\par\end{centering}
\caption{{\it  These are some examples of amputated diagrams that contribute to $\delta \Gamma_{n}$ at two-loops and to $\delta H^{(3)}$ at three-loops. They all have a $Q$-loop (dashed lines) and inside it a $\check{\Phi}$ (solid lines) or a $\check{V}$ (wiggly lines) propagates. They are all related to each other by momentum derivatives (Ward Identities).} }
\label{two-loop}
\end{figure}

In this section, we review and generalize a very simple but important diagrammatic observation that was made in \cite{Pomoni:2011jj}. When drawing the diagrams  that must be computed for the calculation of the $SU(2,1|2)$ Hamiltonian, one discovers \cite{Pomoni:2011jj} that: 

{\large \center \it
For any $\mathcal{N}=2$ superconformal theory the only possible way to make diagrams different from the ${\mathcal{N}=4}$ ones is to make a loop with quarks $Q$ (or $\tilde{Q}$) and  let a  $\check{V}$ or $\check{\Phi}$ propagate inside this loop!}
\\
\\
Examples of such Feynman diagrams are given in Figure \ref{two-loop}.
Note that we are working with {\it the interpolating theory} \eqref{actionN2}. When we consider the $SU(2,1|2)$ sector made out of unchecked fields ($\Phi$, $V$),
 a checked field ($\check{\Phi}$, $\check{V}$) must necessarily propagate inside the $Q$-loop in order to make a  diagram  different from the ${\mathcal{N}=4}$ one, and vice versa.  This is due to the fact that we need a vertex with the second coupling constant $\check{g}$. By power counting, this observation immediately pushes the possibility for  $\delta H = H_{\mathcal{N}=2} - H_{\mathcal{N}=4} \neq 0$ to three-loops. In the next section we  explain this in detail.
We use the language of ${\mathcal{N}=1}$ superspace closely following  \cite{Pomoni:2011jj}. The wiggly lines depict   $\mathcal{N}=1$ vector superfields $V$ (or $\check{V}$) while the solid lines denote   $\mathcal{N}=1$ chiral superfields $\Phi$ (or $\check{\Phi}$). The dashed lines denote bi-fundamental hyper-multiplets $Q$ and $\tilde Q$.

\bigskip

\begin{figure}[h!!]
   \centering
   \includegraphics[scale=0.5]{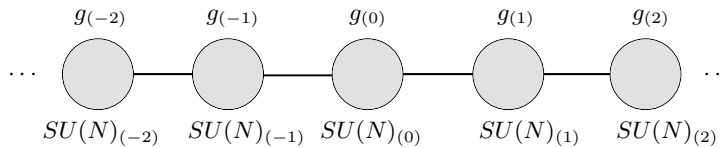}
  \put(-19,35){{\footnotesize $g_{(2)}$}}
    \put(-25,-10){{\footnotesize $SU(N)_{(2)}$}}
      \put(-70,35){{\footnotesize $g_{(1)}$}}
    \put(-77,-10){{\footnotesize $SU(N)_{(1)}$}}
        \put(-126,35){{\footnotesize $g_{(0)}$}}
    \put(-137,-10){{\footnotesize $SU(N)_{(0)}$}}
     \put(-180,35){{\footnotesize $g_{(-1)}$}}
    \put(-187,-10){{\footnotesize $SU(N)_{(-1)}$}}
  \put(-230,35){{\footnotesize $g_{(-2)}$}} 
    \put(-242,-10){{\footnotesize $SU(N)_{(-2)}$}}
        \put(-255,12){{\footnotesize $\cdots$}}
\put(8,12){{\footnotesize $\cdots$}}
 \caption{\it 
 This is a part of a quiver. The vector multiplet in the center denoted by $SU(N)_{(0)}$ has coupling constant $g_{(0)}$ and fields content $\left( V^{(0)} ,\Phi^{(0)}\right)$, then $SU(N)_{(1)}$ and $SU(N)_{(-1)}$ denote the nearest neighbor color group with coupling constants $g_{(1)}$ and  $g_{(-1)}$, respectively. Finally,  $SU(N)_{(2)}$ and $SU(N)_{(-2)}$ denote the next to nearest neighbors with coupling constant $g_{(2)}$ and  $g_{(-2)}$.}
 \label{quiver}
\end{figure}

\begin{figure}[h!!]
   \centering
   \includegraphics[scale=0.7]{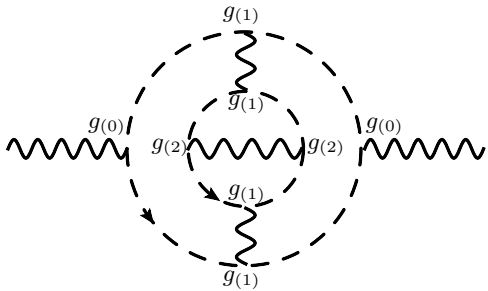}
   \put(-46,53){{\footnotesize $g_{(0)}$}}
      \put(-151,53){{\footnotesize $g_{(0)}$}}
      \put(-68,45){{\footnotesize $g_{(2)}$}}
      \put(-127,45){{\footnotesize $g_{(2)}$}}
 \put(-100,95){{\footnotesize $g_{(1)}$}}
  \put(-100,-5){{\footnotesize $g_{(1)}$}}
   \put(-98,61.5){{\footnotesize $g_{(1)}$}}
  \put(-98, 27.5){{\footnotesize $g_{(1)}$}}
 \caption{\it 
 This ``nested'' Feynman diagram illustrates how to four loops the different coupling constants of the quiver enter in the self-energy corrections of $V^{(0)}$. We begin with $g_{(0)}$ at one loop, then $g_{(1)}$ enters at two loops, $g_{(2)}$ at four etc.
 This means that in the computation of the Hamiltonian $g_{(0)}$ enters at one loop. Then $g_{(1)}$ and  $g_{(-1)}$ appear at tree loops and $g_{(2)}$ and  $g_{(-2)}$ at five loops and so on.
 }
 \label{nested}
\end{figure}

For more general ADE quivers with more than two color groups the corrections will appear as follows in the operator renormalization diagrams.
 At one and two loops only the coupling constant $g_{(0)}$ that corresponds to the vector multiplet of the sector $\left( V^{(0)} ,\Phi^{(0)}\right)$ will appear.
At three loops the
nearest neighbor (in the quiver diagram) color group coupling constant  $g_{(1)}$ (and $g_{(-1)}$) kicks in,
and then the next to nearest neighbor $g_{(2)}$  (and $g_{(-2)}$)  at five loops and so on. See Figures \ref{quiver} and \ref{nested} while remembering that four-loop self-energy corrections enter the calculation of the five-loop Hamiltonian.

\bigskip

\section{One, two and three loops}
\label{123}
In this section we will study and classify the diagrams that need to be computed in order to obtain the difference between the dilatation operator of the $\mathcal{N}=2$ interpolating theory that we presented in section \ref{interpolating} and the dilatation operator of  $\mathcal{N}=4$ SYM 
\be
\delta H = H_{\mathcal{N}=2}- H_{\mathcal{N}=4}
\ee
in the $SU(2,1|2)$ sector, up to three loops, and in the large $N$ limit.
 Almost all the problems that we will face in the next Section \ref{All-loops} with the all-loop argument already appear to  three loops.
  So, it is important to first deal with them here, where we can draw concrete examples of diagrams and discuss their structure and consequences. 
 To be more precise, all the problems concerning single-vertex insertions  appear up to three loops. For multi-vertex insertions one would have to study four-, five-loop  and six- loop examples. This is done in the Appendix \ref{multi-app} where some examples are presented.

\subsection{One-loop}


\begin{figure}[h!!]
   \centering
   \mbox{
   \includegraphics[scale=0.45]{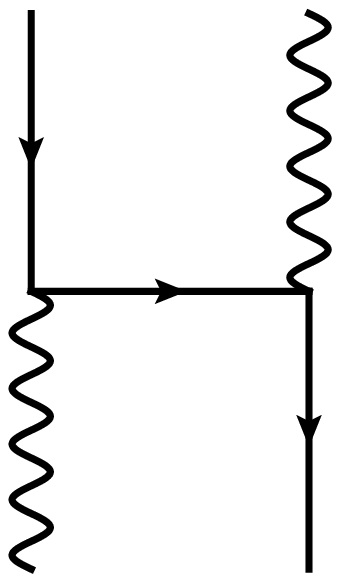}
          \put(-28.5,-15){{\footnotesize ($i$)}}
       \qquad    \qquad    \qquad
          \qquad    \qquad    \qquad
       \includegraphics[scale=0.45]{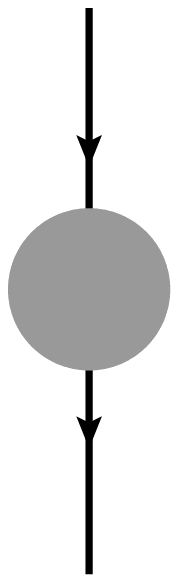}
         \put(-20.5,35){{\tiny 1-$loop$}
          \put(-19.5,-50){{\footnotesize ($ii$)}}
         }
         }
 \caption{\it At one-loop, the only types of diagrams that can contribute to the renormalization of an operator are tree-level skeleton diagrams of order $g^2$ and one-loop self-energy  corrections of the fields that appear in the operator.}
 \label{1-loop}
\end{figure}

%

At one-loop it is very simple to calculate the dilatation operator in the $SU(2,1|2)$ sector of any gauge theory. The only classes of diagrams that could contribute are
\begin{itemize}
\item[($i$)]
The $g^2$ {\it bare skeleton diagrams} that are identical to the ones of $\mathcal{N}=4$ SYM.
\item[($ii$)]
The one-loop self-energy correction of all the fields that appear in the operator. 
\end{itemize}
The  {\it bare skeleton diagrams}  are identical\footnote{This statement should be understood in the following way: the bare skeleton diagrams that are evaluated with the $\mathcal{N}=2$ vertices of the action \eqref{actionN2} give the same result as  the bare skeleton diagrams  evaluated with the $\mathcal{N}=4$ SYM vertices.} to the ones of $\mathcal{N}=4$ SYM for {\it any} gauge theory, because the tree level vertices that make up these diagrams are identical to the $\mathcal{N}=4$ ones.
It is important to realize that for the $SU(2,1|2)$ sector Hamiltonian there are {\it no bare skeleton diagrams made out of vertices coming from the superpotential}!
This is in strict distinction with the $SU(2)$ sector of $\mathcal{N}=4$ SYM presented in \cite{Sieg:2010tz} and the deformed $SU(2)$ sector of the $\mathcal{N}=2$ interpolating theory presented in \cite{Pomoni:2011jj} that is made out of hyper-multiplets.
The one-loop divergencies in the self-energy diagrams are, for any  superconformal gauge theory   \cite{Pomoni:2011jj},   identical to the ones in $\mathcal{N}=4$. In fact, in superspace,  the one-loop divergencies in the self-energy diagrams are immediately equal to zero.
In Figure \ref{1-loop}$(i)$ a representative of the $g^2$ {\it bare skeleton diagrams} is depicted. Therefore, we get
\be
H^{(1)}_{\mathcal{N}=2}(g) = H^{(1)}_{\mathcal{N}=4}(g)  \, .
\ee
A comment for the technically inclined reader is in order.
When we use  ${\mathcal{N}=1}$ superspace, the self-energy correction is  zero due to the fact that the Wess Zumino (WZ) gauge is not fixed. In components, we obtain a non-zero answer \cite{Gadde:2010zi,Liendo:2011xb}, as an artifact of the WZ gauge \cite{Kovacs:1999rd}.

%

\bigskip

\subsection{Two-loops}
\label{Two-loops}
For the two-loop renormalization of operators in the $SU(2,1|2)$ sector one would have to compute the following classes of diagrams:
\begin{itemize}
\item[($i$)]
$g^4$ {\it bare skeletons} 
\item[($ii$)]
{\it dressed skeletons}:  $g^2$ {\it tree-level skeletons}  from the previous one-loop order now {\it dressed} with the insertion of an  one-loop self-energy or one-loop  vertex corrections 
\item[($iii$)]
  the two-loop self-energy corrections
    \item[($iv$)]
{\it new vertices} that are created at one-loop in the effective action \cite{deWit:1996kc} of  $\mathcal{N}=2$ theories $\Gamma^{(1)}_{new}$.
\end{itemize}
Obviously,  the $g^4$ {\it bare skeletons} are identical to the $\mathcal{N}=4$ ones; for skeletons with the same external field content.
In addition, the $g^2$ {\it bare skeletons} that are dressed with one-loop corrections are identical to the $\mathcal{N}=4$ ones {\it iff} the theory is conformal.
   Moreover,   in a conformal theory
 the two-loop self-energy corrections do not contribute to the difference $\delta H$ because their logarithmic divergences are zero.\footnote{In \cite{Pomoni:2011jj} the two-loop self-energy of $\Phi$ was computed. Then, $\mathcal{N}=2$ supersymmetry guarantees that $Z_{W_\alpha} (g)= Z_\Phi(g)$ as we discussed in section \ref{BFF}. 
For an all-loop argument using $\mathcal{N}=2$ superspace see \cite{Howe:1983wj}.
 } 
Thus, the only diagrams that need to be separately discussed are the ones that include {\it new vertices} that appeared in the effective action for the first time at one-loop $\delta\Gamma_{new}^{(1)}$.
\begin{figure}[h!!]
\begin{centering}
\includegraphics[scale=0.35]{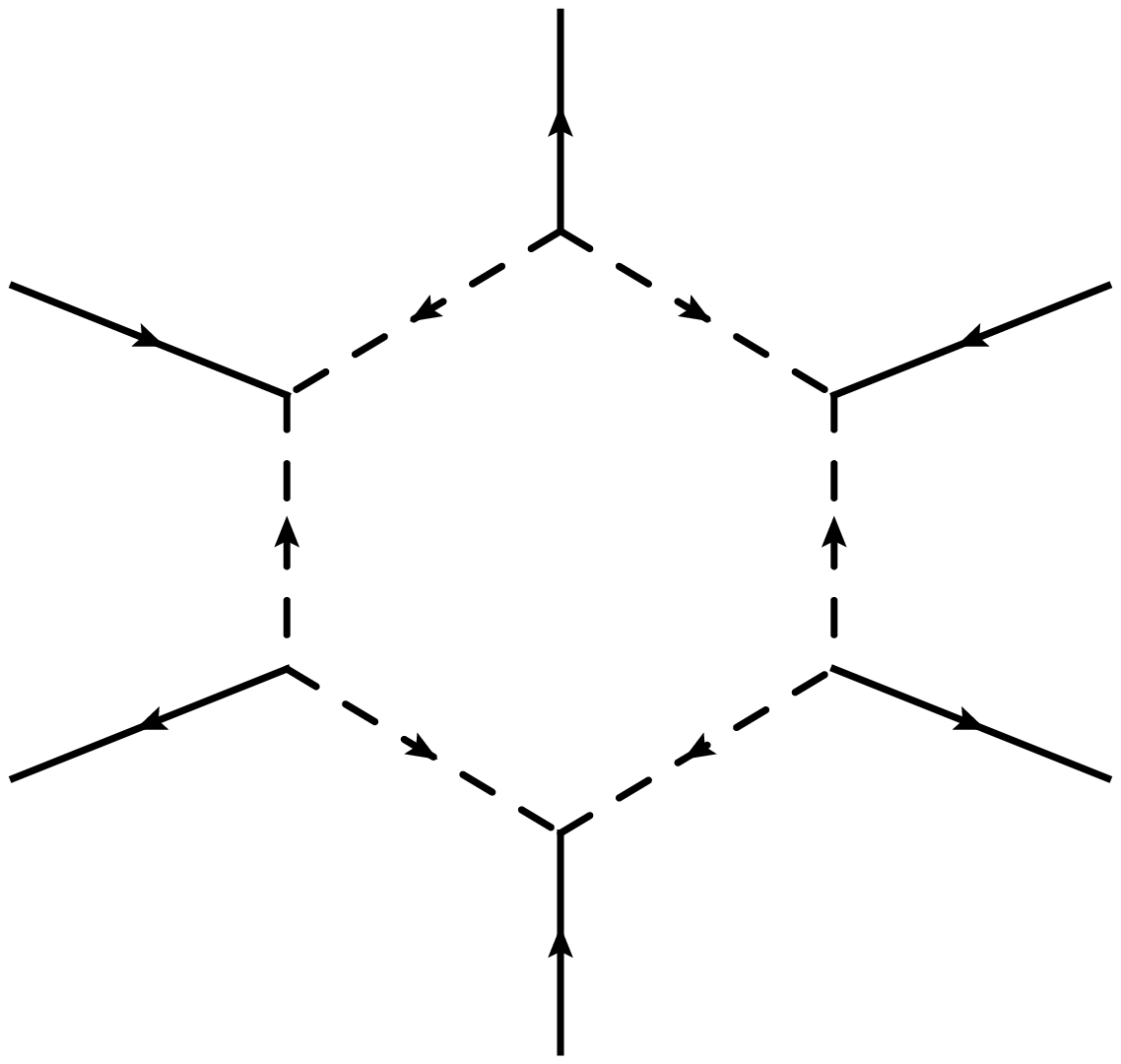}
\qquad \qquad
\includegraphics[scale=0.4]{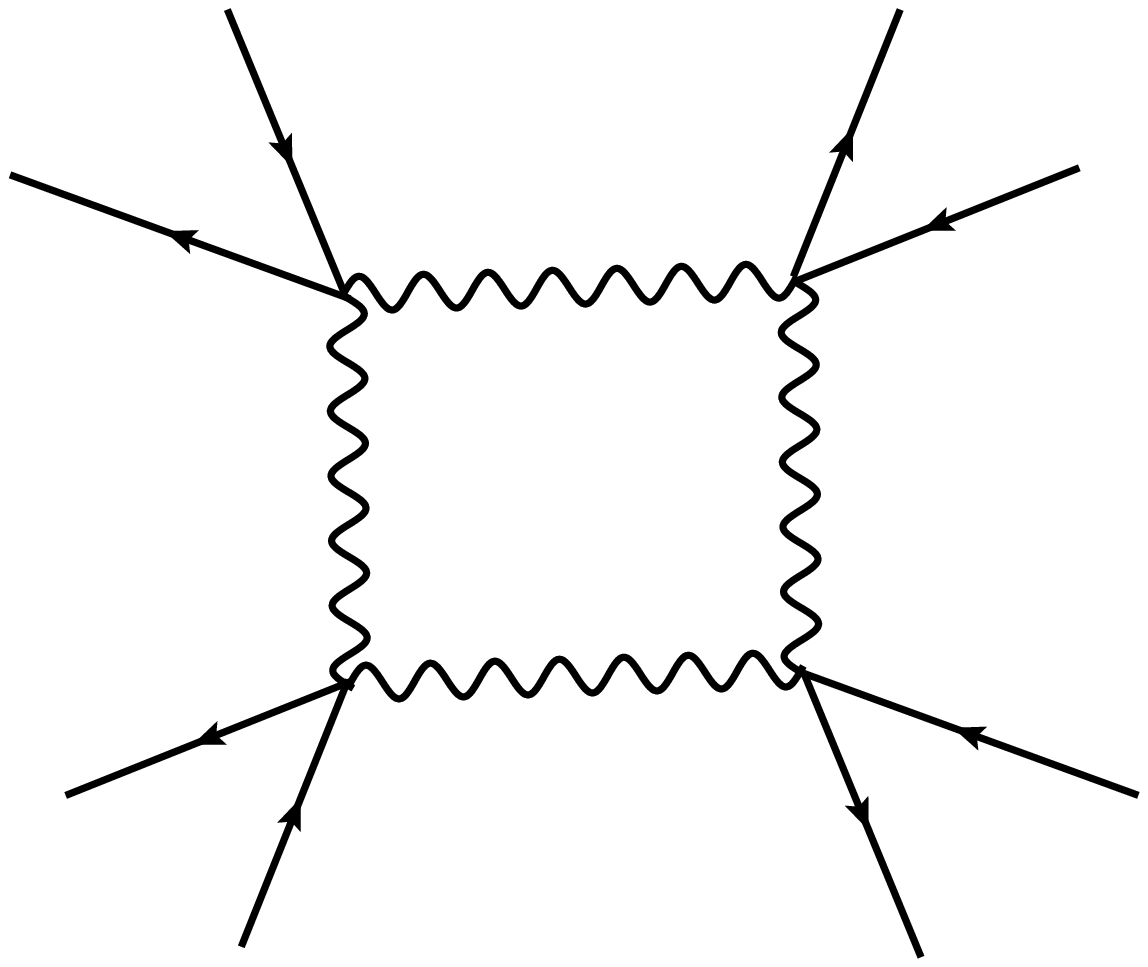}
 \put(-243,-15){{$(a)$}}
 \put(-72,-15){{$(b)$}}
\par\end{centering}
\caption{{\it Examples of the diagrams that are responsible for new vertices in the one-loop effective action of $\mathcal{N}=2$ gauge theories and computed in \cite{deWit:1996kc}. 
A careful examination of this figure tells us that these diagrams cannot be made planar if we require all the fields $\Phi$ to be next to each other.}}
\label{1-loop-new}
\end{figure}

\bigskip

The one-loop corrections to the effective action $\Gamma^{(1)}_{new}$ of an $\mathcal{N}=2$ gauge theory with fundamental hypermultiplets ($Q, \tilde Q$) were computed in \cite{deWit:1996kc}. 
In $\mathcal{N}=2$ superspace language the new terms originate from the real function $\mathcal{H}(\mathcal{W},\bar{\mathcal{W}})$ of the chiral (anti-chiral) scalar superfield strength $\mathcal{W}$ ($\bar{\mathcal{W}}$) which is integrated with the full $\mathcal{N}=2$ superspace measure
\be
S_\mathcal{H} = \int d^4x d^4\theta d^4\tilde \theta \,  \mathcal{H}(\mathcal{W},\bar{\mathcal{W}})   \, ,
\ee
as opposed to the holomorphic prepotential $\int d^2\theta d^2\tilde \theta \, \mathcal{W}^2$ of the tree-level action \eqref{classical-prepotential}.
Diagrams with external $\Phi$ and $\bar\Phi$ are depicted in Figure \ref{1-loop-new}. The diagrams in (a) depict the hypermultiplet  contribution, while those in (b) illustrate the $\mathcal{N}=2$ vector multiplet contribution. The diagrams in (b)  are identical to the $\mathcal{N}=4$ ones for any $\mathcal{N}=2$ gauge theory and give zero when we compute the difference $\delta\Gamma_{new}^{(1)}$. On the other hand, the diagrams in (a) will lead to a non-zero contribution  if the $\mathcal{N}=2$ theory is non-conformal, but  when the number of the hypermultiplets ($Q, \tilde Q$) is such that the the $\mathcal{N}=2$ theory  is conformal  they will be identical to the $\mathcal{N}=4$ ones and finally give
\be
\delta\Gamma_{new}^{(1)}(\Phi,\bar\Phi) =0  
\, .
\ee
Summing everything up, to two loops, we get that the Hamiltonian of a superconformal   $\mathcal{N}=2$ theory  in the $SU(2,1|1)$ sector
\be
H^{(2)}_{\mathcal{N}=2} (g)= H^{(2)}_{\mathcal{N}=4}(g)  \, .
\ee

\bigskip

\subsection{Three-loops}
\label{Three-loops}
In this section, we will discuss which
 three-loop Feynman diagrams appearing in the operator mixing are different from the $\mathcal{N}=4$ ones, in the large $N$ limit.
\begin{figure}[h!!]
   \centering
     \includegraphics[scale=0.45]{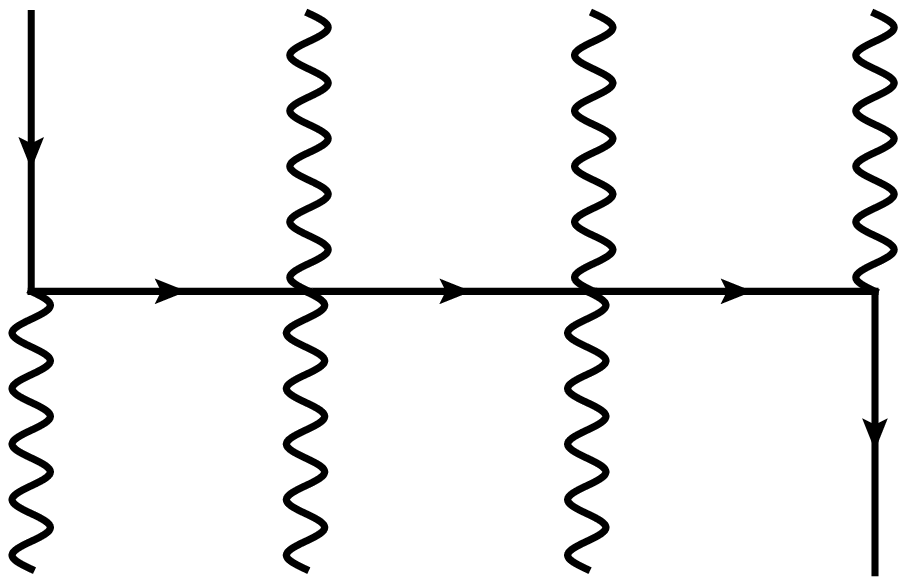}
          \put(-64,-15){{\footnotesize ($i$)}}
        \qquad    \qquad    \qquad  
      \includegraphics[scale=0.45]{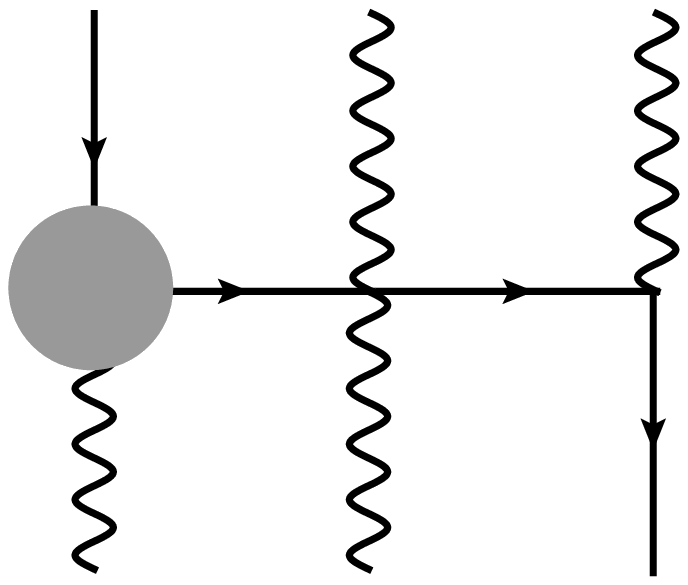}
                \put(-87.5,35.5){{\tiny 1-$loop$}}
          \put(-47,-15){{\footnotesize ($ii$)}}
   \qquad    \qquad    \qquad
   \includegraphics[scale=0.45]{1-loop-renormalized-skeleton.eps}
          \put(-49,35.5){{\tiny 2-$loop$}}
          \put(-28.5,-15){{\footnotesize ($iii$)}}
              \\
              \, ~ \,
\\
\, ~ \,
\\
   \includegraphics[scale=0.45]{self-energy.eps}
          \put(-20,35){{\tiny 3-$loop$}}
          \put(-19.5,-15){{\footnotesize ($iii$)}}
             \qquad    \qquad    \qquad
   \includegraphics[scale=0.45]{1-loop-new.eps}
        \put(-32.5,35){{\tiny 1-$new$}}
          \put(-28.5,-15){{\footnotesize ($v$)}}
          \qquad    \qquad    \qquad
   \includegraphics[scale=0.45]{1-loop-new.eps}
          \put(-32.5,35){{\tiny 2-$new$}}
          \put(-28.5,-15){{\footnotesize ($vi$)}}
 \caption{\it The types of diagrams that contribute to the three-loop dilatation operator are depicted here.  Apart form the bare and the renormalized skeletons we have  the new vertices that appeared at one-loop and are now dressed. Moreover, extra new vertices can appear for the first time at two-loops.}
 \label{3-loop}
\end{figure}

\bigskip

At three loops the diagrams that contribute to the dilatation operator can be classified as follows
\begin{itemize}
\item[($i$)]
$g^6$ {\it bare skeletons} 
\item[($ii$)]
 {\it dressed skeletons} made out of  $g^4$ (or $g^2$) {\it bare skeletons}  {\it dressed} with one  insertion (or two insertions) of  one-loop corrections to their propagators or their vertices  
\item[($iii$)]
 {\it dressed skeletons} made out of $g^2$ {\it bare skeletons}  {\it dressed}  with one insertion of a  two-loop correction to their propagators or their vertices
\item[($iv$)]
 three-loop self-energy corrections
\item[($v$)]
dressed $\Gamma^{(1)}_{new}$  (vertices that appeared for the first time in the  one-loop effective action and are now dressed to become two-loop diagrams)
\item[($vi$)]
new vertices  that appear in the effective action  for the first time at two-loops $\delta \Gamma^{(2)}_{new}$.
\end{itemize}

As before, diagrams that  include $g^6$ {\it bare skeletons} are identical to the $\mathcal{N}=4$ ones for {\it any} gauge theory. For conformal theories, the one-loop corrections with which the $g^4$ or $g^2$  {\it bare skeletons} are dressed are also identical to the $\mathcal{N}=4$ ones, and the three-loop self-energy corrections do not contribute to the difference $\delta H$, as their logarithmic divergence is zero. 
The diagrams that will definitely contribute to the difference $\delta H$ (whether the theory is conformal or not) are the  $g^2$ {\it bare skeletons}  that are {\it dressed}  with one insertion of a  two-loop correction to their propagators and their vertices ({\it dressed skeletons}). 
Following  \cite{Pomoni:2011jj}, the only diagrams that can contribute to $\delta H^{(3)}$ for conformal theories are depicted in Figure \ref{two-loop}. They correspond to two-loop self-energy and vertex corrections. The amputated graphs  depicted in Figure \ref{two-loop} are all related to each other by taking derivatives with respect to momenta (in fact this is how Ward must have discovered the Ward identities).

\bigskip

\begin{figure}[h!!]
   \centering
   \includegraphics[scale=0.5]{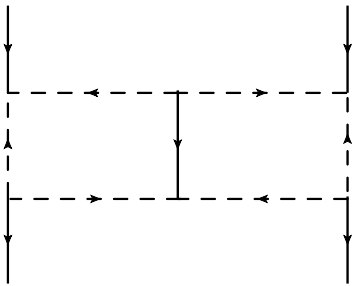} 
 \caption{\it This is the first $\delta \Gamma_{new}$-type diagram that can be planarly Wick-contracted to $\mathcal{O} \in SU(2,1|2)$.  This diagram  is finite \cite{Smirnov:1999gc} and, when  we subtract the $\mathcal{N}=4$ equivalent, leads to a contribution proportional to  $(\check{g}^2 -g^2)g^4$. Due to the non-renormalization theorem of  \cite{Fiamberti:2008sh,Sieg:2010tz}, this diagram gives also a finite contribution when inserted in the operator renormalization diagram, and thus does not contribute to anomalous dimensions. 
 }
 \label{bad-diagram}
\end{figure}

Finally, the {\it new effective vertices} $\delta\Gamma_{new}$ will not contribute to $\delta H$ for the following reasons:
\begin{itemize}
\item
at two loops the effective action will include $\Gamma^{(1)}_{new}$ one-loop vertices from \cite{deWit:1996kc}  that are now dressed by a one-loop insertion or a $V$. These vertices
 {\it do not contribute due to planarity} (they cannot be Wick-contracted with operators of the $SU(2,1|2)$ sector), {\it the choice of the sector} and  {\it Lorentz invariance},  
\item
of the new vertices  that appear in the effective action  for the first time at two loops $\delta \Gamma^{(2)}_{new}$,
the only one that can be planarly contracted to $\mathcal{O} \in SU(2,1|2)$ is shown in Figure \ref{bad-diagram} and  it
 will not contribute due to the non-renormalization theorem of \cite{Fiamberti:2008sh,Sieg:2010tz} that we will further discuss in section \ref{non-renorm-theorem} and in Appendix \ref{6-loop}.
\end{itemize}

\medskip
    
At this point, explaining in detail how {the choice of the sector, planarity and Lorentz invariance} prevent  $\delta\Gamma_{new}^{(1)}(\mathcal{W},\bar{\mathcal{W}})$ from contributing to the Hamiltonian at higher loops when $\check\Phi$ and $\check V$ can propagate inside its loop is in order. The one-loop effective action, as computed in  \cite{deWit:1996kc} by supersymmetrizing the contributions of diagrams such as the ones depicted in Figure \ref{1-loop-new}, is written in that article in equations (22) and (24). Given the choice of our sector \eqref{sector-fields}, the diagrams depicted in Figure \ref{1-loop-new} {\it cannot be planarly contracted to $\mathcal{O} \in SU(2,1|2)$}\footnote{Cannot be planarly contracted at a single trace level. In this paper we are not interested in wrapping corrections. If we wish to include wrapping corrections we will have to take them in account as discussed in 
\cite{Sieg:2005kd}.}. This is because the diagrams in Figure \ref{1-loop-new} include only alternating $\Phi \bar{\Phi} \Phi \bar{\Phi} \cdots$ vertices. Moreover, the supersymmetric completion of $\Phi \bar{\Phi} \Phi \bar{\Phi} \cdots$ includes, for example, vertices of the form  $W^{\alpha}W_\alpha\bar{W}^{\dot{\alpha}}\bar W_{\dot{\alpha}}$,  that cannot possibly be contracted to $\mathcal{O}$ in \eqref{sector-fields} due to the fact that they are {\it Lorentz scalars}, {i.e.} {\it antisymmetric representations of $SU(2)_{\alpha}\times SU(2)_{\dot\alpha}$}, while $\mathcal{O} \in SU(2,1|2)$ is in a {\it symmetric representation of $SU(2)_{\alpha}$}.

  \bigskip
  
 To show this in detail let us consider inserting the  {\it new effective vertex} $W^{\alpha}W_\alpha\bar{W}^{\dot{\alpha}}\bar W_{\dot{\alpha}}$ between $\mathcal{O}(0)$ and $\bar{\mathcal{O}}(x)$
  \be
  \label{example1}
  \langle  \bar{\mathcal{O}}(x)    | \int dy  \, W^{\alpha}W_\alpha\bar{W}^{\dot{\alpha}}\bar W_{\dot{\alpha}} (y)  \, | \mathcal{O}(0)\rangle \, .
  \ee
  To show that this is zero it is enough to perform half of the Wick-contractions. $\mathcal{O}(0)$ must include at least two nearest neighbor ${W}_{+}$ (if it doesn't, we get zero in the large $N$ limit). Wick-contracting,
  \be
    \label{example2}
  \bar{W}^{\dot{\alpha}}\bar W_{\dot{\alpha}} (y)   \, \mathcal{O}(0) \sim   \epsilon^{\dot{\alpha} \dot{\beta}} \bar{W}_{\dot{\alpha}}\bar W_{\dot{\beta}} (y)  \left(   \cdots {W}_{+}{W}_{+}(0)   \cdots \right)\sim 
  \cdots \epsilon^{\dot{\alpha} \dot{\beta}} \,  \frac{y_{+ \dot{\alpha}}}{y^4}   \frac{y_{+ \dot{\beta}}}{y^4} \cdots  =0 \, ,
  \ee
given the fact $\epsilon^{\dot{\alpha} \dot{\beta}}$ is antisymmetric, that the coordinates $y$ are bosonic and thus $y_{+ \dot{\alpha}} y_{+ \dot{\beta}}$ is symmetric.

\medskip

\bigskip

We thus conclude that for conformal $\mathcal{N}=2$ theories the only diagrams that can contribute to $\delta H^{(3)}$ are
the  two-loop corrections of 
the $g^2$ bare skeleton diagrams. This is the third diagram in Figure  \ref{3-loop} with the two-loop (self-energy or vertex) corrections  depicted in Figure \ref{two-loop} inserted in all possible positions.
This is enough to show that
\be
\delta H^{(3)}(g)  = c_3( g)  \,H^{(1)}_{\mathcal{N}=4}  \, ,
\ee
because the external fields structure of the dressed diagram is identical to the structure of the bare diagram. In the case of {\it the interpolating theory}, the only difference between
 dressed and  bare diagrams is a factor
\be
c_3( g)  = c_3( g, \check{g}) \sim c_3 \, g^2\left( g^2 - \check{g}^2\right) \, ,
\ee
where the coefficient $c_3$ includes combinatorial information together with the knowledge of the momentum integrals that are performed.
The coefficient  $c_3( g)$ can be obtained by an explicit calculation, but this is not our goal here.
We just want to notice that it contains all the loop information encoded in $Z(g)$ and the combinatorial information that is obtained by going from the effective action $\Gamma$ (1PI generating functional) to 1PI n-point functions $\Gamma_n$ and finally to connected graphs $G_{c\,n}$. This is why we set up our all-loop argument in terms of the effective action $\delta \Gamma$ and the $Z(g)$.

\medskip

Collecting all the above results, we see that the Hamiltonian up to three loops can be written as
\be
\delta H^{(3)}(g)  = H^{(3)}_{\mathcal{N}=4} (\newg) \, ,
\ee
where $\newg^2=f(g^2,\check{g}^2)$ is a function of $g$ and $\check{g}$ that we can obtain pertubatively.

\bigskip

One last explanation is in order: why does the new vertex that comes from the diagram in Figure \ref{bad-diagram} not contribute to the Hamiltonian when it is Wick-contracted with operators with derivatives?
On the one hand, contracting this  new vertex to an operator  $\mathcal{O} \in SU(2,1|2)$  without derivatives leads to a finite integral  of the form
\be
\int dq  \, \mathcal{I}\left(q^2\right)  \, ,
\ee
where    the integrand $\mathcal{I}\left(q^2\right)$ is a scalar under Lorentz transformations.
On the other hand, contracting it to an operator  $\mathcal{O} \in SU(2,1|2)$  with derivatives leads to an integral the form 
\be
\int dq  \, \mathcal{I}\left(q^2\right) q^{+ \dot\alpha} q^{+ \dot\beta} \cdots  \, .
\ee
{\it Due to Lorentz covariance}, this integral can  either be zero or after partial integration become proportional to
\be
\left( q_{ext_1}^{+ \dot\alpha} q_{ext_2}^{+ \dot\beta}  \cdots  \right)\int dq \, \mathcal{I} \left(q^2\right) \, ,
\ee
where the numerator  momenta with open Lorentz indices  will have to end up outside. The momenta $q_{ext_1}^{+ \dot\alpha}$, $q_{ext_2}^{+ \dot\beta}$, \dots are external to the loop we are integrating over. This integral that we end up with is again finite and will not contribute to the Hamiltonian. In other words,  operators with derivatives in the $SU(2,1|2)$ sector create
 traceless symmetric products of momenta in the numerator of the loop integral that cannot change the divergence structure of the integral.  
 The argument we just gave is for partial derivatives $\partial_{+\dot+}$; not for covariant derivatives. Up to three loops, it is trivial to consider the gauge boson emission diagrams and see that they obey \eqref{statement}. 
 We skip this step because first of all it is very simple and secondly because one should not do it.
 Our logic is that we use of the background field formalism that guarantees gauge invariance. Whatever holds for partial derivatives $\partial_{+\dot+}$ will also hold for covariant derivatives.

\bigskip

\subsection{Length-changing operations}
The educated reader is most probably thinking that this is not all. Up to now we discussed one, two and tree loops, but there are elements of the Hamiltonian that come in between and correspond to  length-changing operations.
\begin{figure}[h!]
\begin{centering}
\includegraphics[scale=0.5]{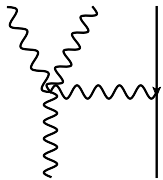}
\par\end{centering}
\caption{{\it Length changing operation at order $g^3$. This diagram is a bare skeleton.}}
\label{length-changing-1loop}
\end{figure}
Let's first consider the first length-changing operations  appearing  at order $g^3$, which we like to call ``{\it one-loop and a half} '', or $H^{(1.5)}$. An example of such an operation is depicted in Figure \ref{length-changing-1loop}.
Such a diagram is a {\it bare skeleton}, and as such it is identical to the  $\mathcal{N}=4$ one.

\medskip

\begin{figure}[h!]
\begin{centering}
\includegraphics[scale=0.5]{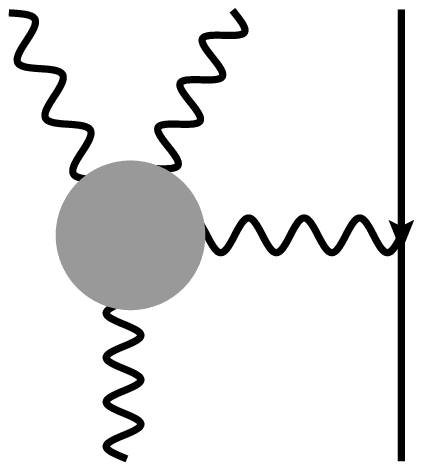}
\put(-51,31.5){\tiny 1-$loop$}
\qquad  \qquad \qquad
\includegraphics[scale=0.5]{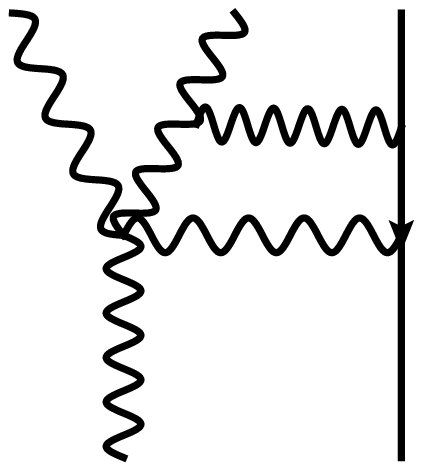}
\par\end{centering}
\caption{{\it Examples of length changing operations at at order $g^5$.}}
\label{length-changing-2loop}
\end{figure}

At order $g^5$ is when the next length-changing operation can occur ($H^{(2.5)}$ or ``{\it two-loops and a half} ''). Examples of such an operation are depicted in Figure \ref{length-changing-2loop}.
It is clear that the only thing one can do to the diagram is to either correct the  $g^3$ length-changing diagram by a one-loop vertex (first example) or by a one-loop self-energy correction, or attach an extra gluon (second example). As we have discussed above all these possibilities cannot make the Hamiltonian different from the $\mathcal{N}=4$ one.

\medskip

For conformal $\mathcal{N}=2$ theories, only starting at order $g^7$ (``{\it three-loops and a half} '') we can  have length-changing diagrams different from the  $\mathcal{N}=4$ ones by inserting in
 the diagram of Figure \ref{length-changing-1loop} corrections of the form of Figure
\ref{two-loop}. However, this diagram is a {\it dressed skeleton} that will lead to $\delta H^{(3.5)} \sim g^2(g^2-\check{g}^2) H^{(1.5)}_{\mathcal{N}=4}$ up to a combinatorial factor and thus will obey \eqref{statement}.

\section{All loops}
\label{All-loops}

At this point a clear pattern is emerging. 
\\
{\large  \it
The only non-zero contribution  to the difference of the Hamiltonians, 
\be
\delta H = H_{\mathcal{N}=2} -  H_{\mathcal{N}=4}\, ,
\ee
 in the $SU(2,1|2)$ sector is due to the different dressings with $\delta Z(g)$ of the bare skeleton diagrams.
}
\\
 In principle, one should also consider {\it new effective vertices that will appear in the effective action} at some loop order, but as we will see in this section {\it these new vertices can never contribute to the Hamiltonian of the $SU(2,1|2)$ sector}!
Then, given the fact that the {\it dressed skeletons} have precisely the same structure as the {\it bare skeletons},  the $\ell$-loop Hamiltonian is corrected by 
\be
\delta H^{(\ell)} \sim \sum_{\ell' =1}^{\ell-2} c^{\ell}_{\ell'} (g)H^{(\ell')}_{\mathcal{N}=4} \, .
\ee
The coefficients $c^{\ell}_{\ell'} (g)$  include two different pieces of information. The first piece of information is the combinatorial factors that
one obtains from the computation of the  connected graphs $G_{c\,n}$ starting from the 1PI generating functional $\Gamma$ and, finally, the Wick-contractions with $\mathcal{O}$ and $\bar{\mathcal{O}}$. 
The second one is the dynamics, the effects of the loops and renormalization, and is encoded in a single function $\delta Z(g)$. The combinatorial factors are the same as in $\mathcal{N}=4$ exactly because the {bare skeletons} with external fields only in the vector multiplet of any gauge theory are identical to the $\mathcal{N}=4$  ones.
However, the $Z(g)$ for a particular $\mathcal{N}=2$ superconformal theory is of course different from $\mathcal{N}=4$ and $\delta Z(g)$ leads to the unique and universal function $f(g^2)$ that encodes the redefinition of the coupling constant
$ g^2 \rightarrow f(g^2)$.

\bigskip

 All this information is elegantly encoded  in
 the effective action without the difficulty of having to keep track of combinatorial factors. Our strategy is  to consider, at any loop order, the difference in the two effective actions $\delta \Gamma =  \Gamma_{\mathcal{N}=2} -  \Gamma_{\mathcal{N}=4}$, which we will think of as the sum 
\be
\delta \Gamma = \delta \Gamma_{ren. \, tree} + \delta  \Gamma_{new}
\ee
of terms that  were already there at the tree level and are now renormalized $\delta  \Gamma_{ren. \, tree}$ and vertices that were not there at {tree level} $\delta  \Gamma_{new}$.

\bigskip

In the next two sections we discuss the structure of possible new vertices  in the effective action. Most of the terms in $\delta  \Gamma_{new}$ cannot contribute to the Hamiltonian of the $SU(2,1|2)$ sector because they cannot be Wick-contracted with an operator $\mathcal{O}\in SU(2,1|2)$ due to {\it planarity, Lorentz invariance of $\delta  \Gamma_{new}$ and the choice of the sector}.
The ones that in principle could contribute (given in \eqref{new-can-N2supersp})
 do not lead to logarithmic divergencies due to a non-renormalization theorem, described in section \ref{non-renorm-theorem}.

\bigskip

From this moment on we stop using $\mathcal{N}=1$ superspace language and turn to $\mathcal{N}=2$ superspace.
We do this because in $\mathcal{N}=2$ superspace all the fields in the $\mathcal{N}=2$ vector multiplet are packed in a single $\mathcal{N}=2$ superfield $\mathcal{W}$.
We also want to stress that the Feynman diagrams that we draw in this section are also in $\mathcal{N}=2$ superspace  and solid lines now depict  $\mathcal{W}$.

\bigskip

\subsection{Classification of possible new vertices in the effective action}
\label{classification}

For {\it conformal} $\mathcal{N}=2$  theories {\it the possible new terms that can appear in the effective action}  have been extensively studied  \cite{deWit:1996kc,Lindstrom:1996xi,Dine:1997nq,Buchbinder:1999jn,Kuzenko:2003wu} and {\it classified} in \cite{Buchbinder:1999jn}
by studying all the possible superconformal invariants. Schematically, they are\footnote{The effective action as written in \eqref{effective} is actually only for the case where we are in the Coulomb branch and $SU(N)$ is broken down to $U(1)^{N-1}$. However, we just write this to avoid cluttering the notation. For a non-abelian version one can see for example \cite{Banin:2002mf,Buchbinder:1998qd}.
}
\bea
\label{effective}
&
\Gamma_{new}(\mathcal{W}) = \int{d^4x \, d^8\theta \, \mathcal{H}(\mathcal{W},\bar{\mathcal{W}})  }
&
\\ \nonumber
& 
 \mathcal{H}(\mathcal{W},\bar{\mathcal{W}}) = \ln^2\left(  \mathcal{W}  \bar{\mathcal{W}} \right)
+\left[  \Lambda (\bar\Psi^2)  \ln \mathcal{W} + h.c. \right]
 + \Upsilon (\Psi^2 , \bar\Psi^2)  + F(\Psi^2 , \bar\Psi^2)
 &
\eea
where $\Lambda$ and $\Upsilon$ are arbitrary holomorphic and real analytic functions, respectively, of
\be
\label{psi's}
\Psi^2 = \frac{1}{\mathcal{W}^2}\bar D^4\ln\bar{\mathcal{W} }
\, ,  \quad \bar{\Psi}^2 = \frac{1}{\bar{\mathcal{W}}^2}  D^4\ln \mathcal{W}
\quad \mbox{with} \quad D^4 = (D^{\mathcal{I}=1})^2 (D^{\mathcal{I}=2})^2    \, ,
\ee
while $F$ is a function\footnote{ More information on the form of the function $F$ can be found in equations (2.14) and (2.15) of \cite{Buchbinder:1999jn}.}   of $\Psi^2$, $\bar{\Psi}^2$ and the derivatives combination $D^{\mathcal{I J}} = D^{\alpha \, \left(\mathcal{I} \right.  } D^{\left. \mathcal{J}\right)}_{\, \alpha}$.

\medskip

 Due to {\it planarity, Lorentz invariance} of  $\delta  \Gamma_{new}$ and the {\it choice of the $SU(2,1|2)$ sector} most of these terms can immediately be excluded.
From \eqref{effective}, {\it the only other possible terms that can contribute} to anomalous dimensions in the $SU(2,1|2)$ sector have the form $\tr \left( \mathcal{W}^n \bar{\mathcal{W}}^n\right)$. 
To see this we firstly notice that vertices that include alternating $\tr\left( \mathcal{W}\bar{\mathcal{W}} \mathcal{W} \bar{\mathcal{W}} \cdots\right)$ cannot be contracted to operators of the $SU(2,1|2)$ sector.
This is precisely the same argument as the one we used in Section \ref{Two-loops}.

 \medskip

 This means that {\it the only} new vertices in the effective action that {\it can} in principle {\it contribute to the $SU(2,1|2)$ sector} have the form
 \be 
  \label{new-can-N2supersp}
\delta  \Gamma_{new}^{can}   =\sum_n  c_n \tr \left( \mathcal{W}^n \bar{\mathcal{W}}^n\right) \, .
\ee
We use the notation $\delta  \Gamma_{new}^{can}$ to remind us that this is a subset of the $\delta  \Gamma_{new}$ vertices that can in principle be Wick-contracted to the operator $\mathcal{O}$ in the $SU(2,1|2)$ sector.
In $\mathcal{N}=1$ superspace language these vertices include
$
\tr \left( \mathcal{W}^n \bar{\mathcal{W}}^n\right)|_{\tilde{\theta}=0}  = \tr  \left( \Phi^n \bar{\Phi}^n\right) + \dots
$. They can in principle lead to elements in the Hamiltonian that are proportional to the chiral identity,
but, as we will discuss in the next section, they do not contribute logarithmic divergencies due to the non-renormalization theorem of \cite{Fiamberti:2008sh,Sieg:2010tz}.

\bigskip

\subsection{A non-renormalization theorem}
\label{non-renorm-theorem}

The  non-renormalization theorem of \cite{Fiamberti:2008sh,Sieg:2010tz} was proved using $\mathcal{N}=1$ superspace formalism and
is based on powercounting and the structural properties of Feynman diagrams in $\mathcal{N}=1$ superspace.
For us, the important lesson is that insertions of the form $\langle  \Phi^n (x) \bar{\Phi}^n(y)  \rangle$, which are {\it proportional to the chiral identity}\footnote{The authors of \cite{Fiamberti:2008sh,Sieg:2010tz} sometimes refer to  the chiral identity as the trivial chiral
function $\chi()$ with no argument, to remind us that nothing is permuted.
 They state that finiteness conditions imply 
that diagrams with trivial chiral
function $\chi()$ cannot have an overall UV divergence.
},  do not contribute to the renormalization of operators.
 An example of such an insertion is depicted in Figure \ref{bad-diagram} and in the Appendix \ref{non-ren-ex} we show by powercounting that it will not lead to UV divergence once it is inserted in the operator renormalization diagram.
This result is also true for many such insertions in the operator renormalization diagram as long as the final connected graph has structure proportional to the chiral identity. 
An example of a connected graph made out of two vertices is also worked out in the Appendix \ref{non-ren-ex}.
 This non-renormalization theorem reflects the fact that chiral operators $\mathcal{O}_L(\Phi) = \tr \left( \Phi^L\right)$ are protected as members of the chiral ring.
 Even though this theorem was derived in $\mathcal{N}=1$ superspace,  one can easily reformulate it in $\mathcal{N}=2$ language, to reflect the fact that chiral operators $\mathcal{O}_L(\mathcal{W}) = \tr \left( \mathcal{W}^L\right)$ are also protected, since they are members of the $\mathcal{N}=2$ chiral ring.

\medskip

The derivation of the theorem is very technical and we will skip it here.
The interested reader is invited to read  \cite{Fiamberti:2008sh,Sieg:2010tz} for the $\mathcal{N}=1$ superspace proof. Below we just present the main points that one would have to change when going from  $\mathcal{N}=1$ to $\mathcal{N}=2$ superspace in order to rederive the theorem in $\mathcal{N}=2$ language. One would also need to write down the Feynman rules explicitly in order to do the power counting. For that the naive real superspace $\mathbb{R}^{4|8}$ that we use here might be way too complicated \cite{Howe:1982tm} and one should maybe instead turn to
the $\mathcal{N}=2$ Harmonic \cite{Galperin:2001uw} or Projective \cite{Lindstrom:1989ne,Jain:2010gm,Jain:2013hua} superspace formalism.

\medskip

The $2$-point functions of the form $\langle  \Phi^n (x) \bar{\Phi}^n(y)  \rangle$ with $n>1$ come with non-negative powers of $\varepsilon$ in dimensional regularization. Moreover, when they are inserted in the operator renormalization diagrams, the diagrams remain finite or less. 
In usual (non-chiral) operator renormalization diagrams when a finite vertex is inserted, 
an   overall $1/\varepsilon$ divergence is obtained. However, a chiral loop cannot create a $1/\varepsilon$ contribution, due to the following facts: ({\it i}) that four $D$'s (superspace derivatives) are used from the numerator each time we perform a $\theta$ integral ($\int d^4\theta$) \cite{Gates:1983nr}, ({\it ii})  that the chiral operator comes with one less $\bar{D}^2$ \cite{Fiamberti:2008sh,Sieg:2010tz}\footnote{
The chiral superfields obey the constraint 
$\bar{D}_{\dot\alpha}\Phi = 0$
that we
resolve using an unconstrainted superfied $\Phi  =\bar{D}^2\varphi$ before doing any Feynman diagram computation.
To obtain the two point function $\langle \bar{\mathcal{O}} \mathcal{O} \rangle$ we have to add to the path integral  a source term 
$\int d^2 \theta  \, j \, \mathcal{O}$.
In order to complete its integral $\int d^2 \theta \rightarrow \int d^4 \theta$ we steal a $\bar{D}^2$ from the operator
$\mathcal{O} = \tr \left( \Phi^L \right)   = \tr \left( \bar{D}^{2L}\varphi^L \right)$
and thus each chiral operator comes in the Feynman diagram with one less  $\bar{D}^2$. Unconstrained $\mathcal{N}=2$ superfiels where introduced in \cite{Howe:1982tm} and further studied and used in \cite{Howe:1983sr}.
}
and  ({\it iii}) a $\bar{D}^2$ can always be moved outside the diagram on the external lines (onto a scalar propagator which is not part of a loop). Thus,
 the ``{\it effective number of $D$'s}''  in the loops is equal to the number of $D$'s minus twice the number of scalar propagators not belonging to any loop \cite{Fiamberti:2008sh}. At the end there are not enough momenta left in the numerator to make the loop divergent.

\smallskip

Similarly, the $2$-point functions of the form $\langle \mathcal{W}^n (x) \bar{\mathcal{W}}^n(y)  \rangle$ do not lead to divergencies when inserted in the operator renormalization diagrams of chiral operators. As in  the $\mathcal{N}=1$ superspace language, a chiral loop cannot create a $1/\varepsilon$ contribution due to the following facts: ({\it i}) that eight superspace derivatives (four $D$'s and four $\tilde{D}$'s) are used from the numerator each time we perform a $d^4\theta d^4\tilde{\theta}$ integral \cite{Jain:2013hua}, ({\it ii})  that the chiral operator comes with one less $\bar{D}^2\bar{\tilde{D}}^2$
and  ({\it iii}) $\bar{D}^2\bar{\tilde{D}}^2$'s can always be moved outside the diagram on external scalar propagators  reducing
 the ``{\it effective number of $D$'s}''  in the loops by four times the number of scalar propagators not belonging to any loop. 
 This means again that there are not enough momenta left in the numerator to make the loop divergent. 

\bigskip

With the use of this non-renormalization theorem we conclude that the new effective vertices \eqref{new-can-N2supersp} appearing in $\delta \Gamma_{new}$ cannot contribute to the Hamiltonian of our sector. 
\be
\langle \bar{\mathcal{O}} (x) | \delta \Gamma^{new}_{n}(y) |  \mathcal{O} (0)\rangle = finite \, .
\ee
For an explicit demonstration see Appendix \ref{non-ren-ex}.
Moreover, this result can also be generalized to the case where a collection of vertices \eqref{new-can-N2supersp} is inserted in the operator  renormalization diagram as long as the final connected graph is proportional to the chiral identity. This statement, together with a few more observations that we will make in section \ref{1PI2connected},
 allow us to conclude that
\be
\langle \bar{\mathcal{O}} (x) | \delta G^{new}_{n}(y) |  \mathcal{O} (0)\rangle = finite \, 
\ee
for every possible connected graph that includes one or more new effective vertex.
The next step is to consider what happens to the  $\delta \Gamma_{ren. \, tree}$ vertices that create the dressed skeleton diagrams.

\bigskip

\subsection{First without derivatives (only skeleton diagrams)}
\label{only-skeletons}

Given the fact that {\it new effective vertices  do not contribute} to the two-point functions $\langle \bar{\mathcal{O}}\mathcal{O}\rangle$, to obtain the Hamiltonian we just have to compute corrections to the propagators, and the $n$-point vertices, that where already there in the Lagrangian at tree level,  and then insert them in the two point functions $\langle \bar{\mathcal{O}}\mathcal{O}\rangle$.
As already discuss in section \ref{BFF}, in the BFM all the information about loops will be encoded in a single function $Z(g)$ (because  $Z_{\mathcal{W}}(g)$ and $Z_{g}(g)$ are related through a WI that reflects gauge invariance).
We first consider the diagrams that lead to elements in the Hamiltonian {\it without any derivatives}.
These are  {\it dressed skeletons} and are encoded in $\delta \Gamma_{ren. \, tree}$.  
Simply by using
\begin{enumerate}
\item Gauge invariance (that is manifest in the background field method)
\item ${\mathcal{N}=2}$ supersymmetry (use ${\mathcal{N}=2}$ superspace)
\end{enumerate}
we get that the tree level action $S (\mathcal{W};g)$ \eqref{classical-prepotential} will be corrected only by a function that is proportional to  the tree level action itself
\be
\label{all-niceL}
\delta  \Gamma_{ren. \, tree}(\mathcal{W};g)   =  \Gamma_{tree} (\mathcal{W};\newg)     \equiv S (\mathcal{W};\newg)     
\ee
where $\newg = \sqrt{f(g^2)}$ is some function of the coupling constant.
This fact can also be understood using $\mathcal{N}=2$ superconformal representation theory considerations. 
With this information at hand, the $n$-point diagrams immediately obey
\be
\label{all-nicegamma-n}
\delta  \Gamma^{ren. \, tree}_{n}(\mathcal{W};g)   =  \Gamma^{tree}_{n} (\mathcal{W};\newg)     
\quad \forall n    \,.
\ee

\subsection{From 1PI to connected graphs}
\label{1PI2connected}

Up to now we have argued using planarity, Lorentz invariance and a non-renormalization theorem that {\it a single insertion} of a {\it new effective vertex} (1PI $n$-point graph) between $\mathcal{O}$ and $\bar{\mathcal{O}}$ is finite
\be
\langle \bar{\mathcal{O}} (x) | \delta \Gamma^{new}_{n}(y) |  \mathcal{O} (0)\rangle = finite  \qquad \forall \,  \mathcal{O} \in SU(2,1|2)  \, ,
\ee
and does not contribute to the Hamiltonian.
Of course this in not enough. For \eqref{statement} to hold we need to argue that the same statement is true not only for 1PI $n$-point graphs, but also for the connected graphs
\be
\langle \bar{\mathcal{O}} (x) | \delta G^{new}_{c\,n}(y) |  \mathcal{O} (0)\rangle = finite  \qquad \forall \,  \mathcal{O} \in SU(2,1|2) \, .
\ee
This will allow us to go from the statement
\be
\delta  \Gamma^{ren. \, tree}_{n}(\mathcal{W};g)   =  \Gamma^{tree}_{n} (\mathcal{W};\newg)     
\quad \forall n    
\ee
that we argued above to
\be
\label{all-nice}
\delta  G_{c\, n}^{contribute}(\mathcal{W};g)   =  G^{tree}_{c\,n} (\mathcal{W};\newg) 
\quad \forall n 
\, .
\ee
In this section, we shall do precisely that, namely provide evidence
 that one can obtain
 the connected off-shell $n$-point functions $\delta G_{c \,n}(\mathcal{W})$ that are relevant for the $SU(2,1|2)$ sector from  the  $n$-point 1PI functions $\delta \Gamma_n(\mathcal{W})$ \eqref{Gamma-n} alone.
This is a highly non-trivial statement and should be further studied by carefully  checking as many explicit examples as possible. 
Some are presented in the Appendix \ref{multi-app}.
This way one might get inspired and manage to formulate and prove this statement with a formal, path integral based argument.
We leave this for future work.

\medskip

Our argument is built on the following simple, but  important observation.
To make a connected  graph   $\delta G_{c \,n}(\mathcal{W})$ that is {\it not} 1PI we must be able to disjoin the graph by cutting a single line.  Inversely, if by cutting an internal propagator we cannot disjoint the graph, then this graph must be a 1PI $\delta \Gamma_{n}(\mathcal{W})$ that we should be able to obtain by taking functional derivatives of the effective action $\delta \Gamma(\mathcal{W})$ with the appropriate number of fields $\mathcal{W}$ and $\bar{\mathcal{W}}$. This means that it corresponds to (or can also be thought of  as) a {\it single new effective vertex}, and we have already explained why {\it a single new effective vertex $\delta \Gamma^{new}_{n}$ cannot contribute to the Hamiltonian}.
  \begin{figure}[h!!]
   \centering
   \includegraphics[scale=0.4]{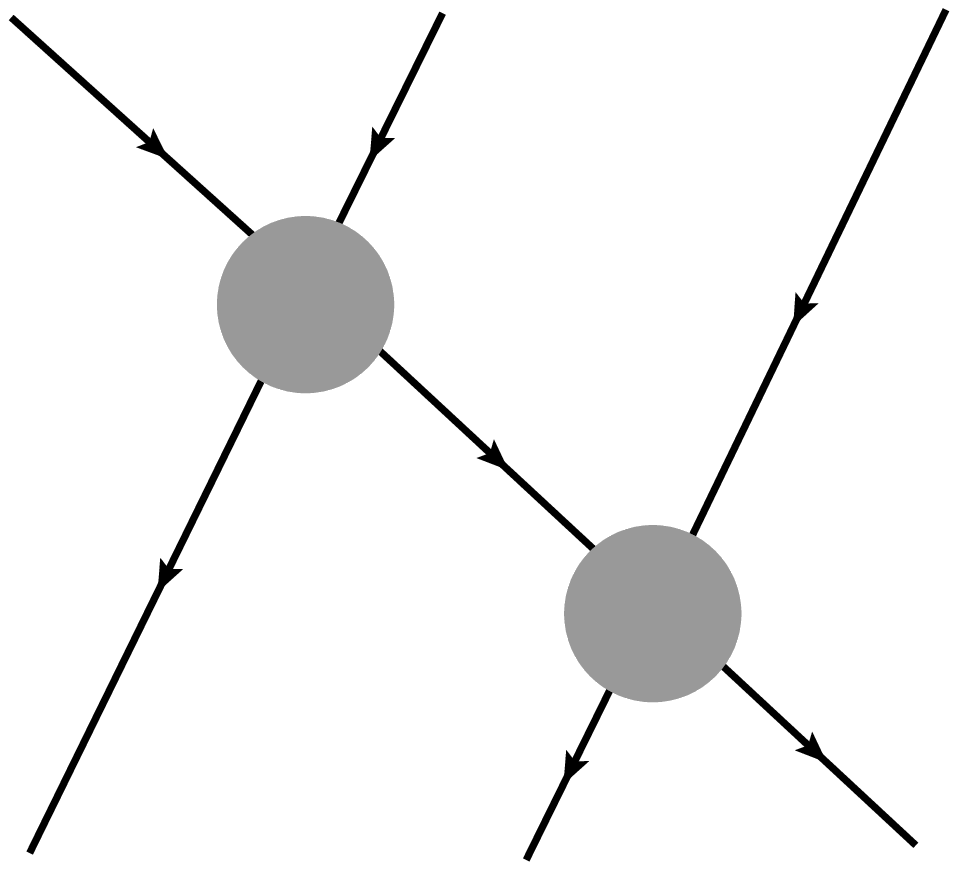}
     \qquad    \qquad   
     \includegraphics[scale=0.4]{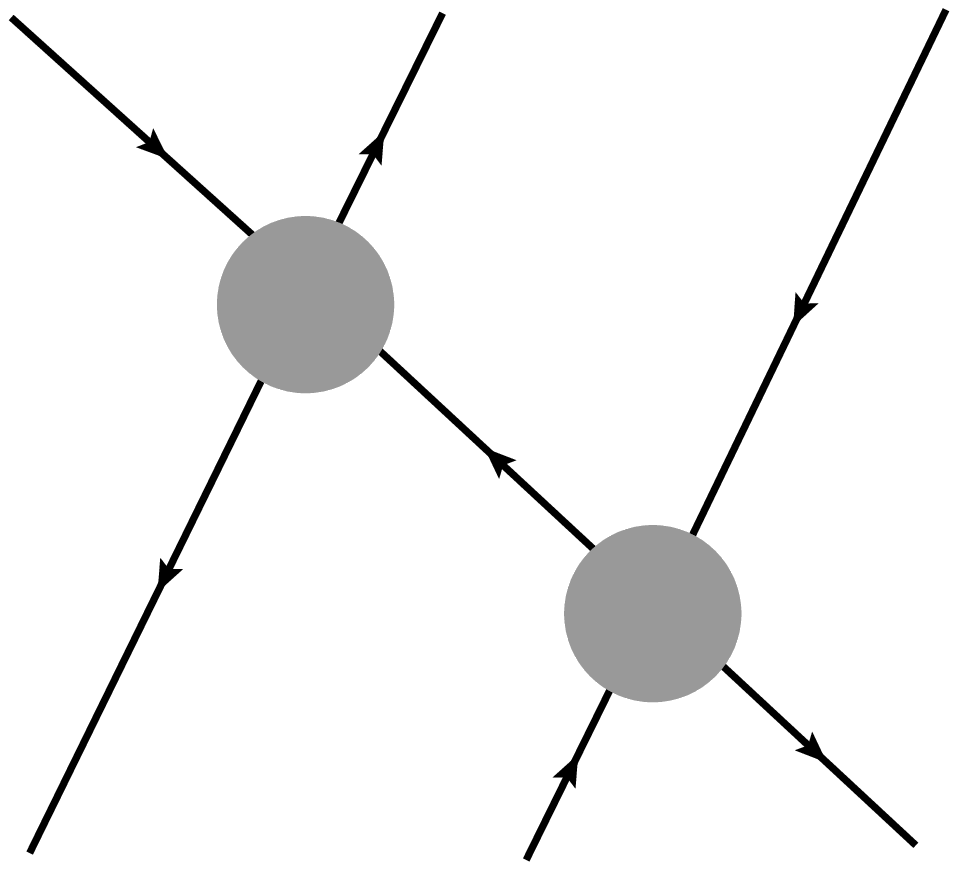}
        \qquad    \qquad   
     \includegraphics[scale=0.4]{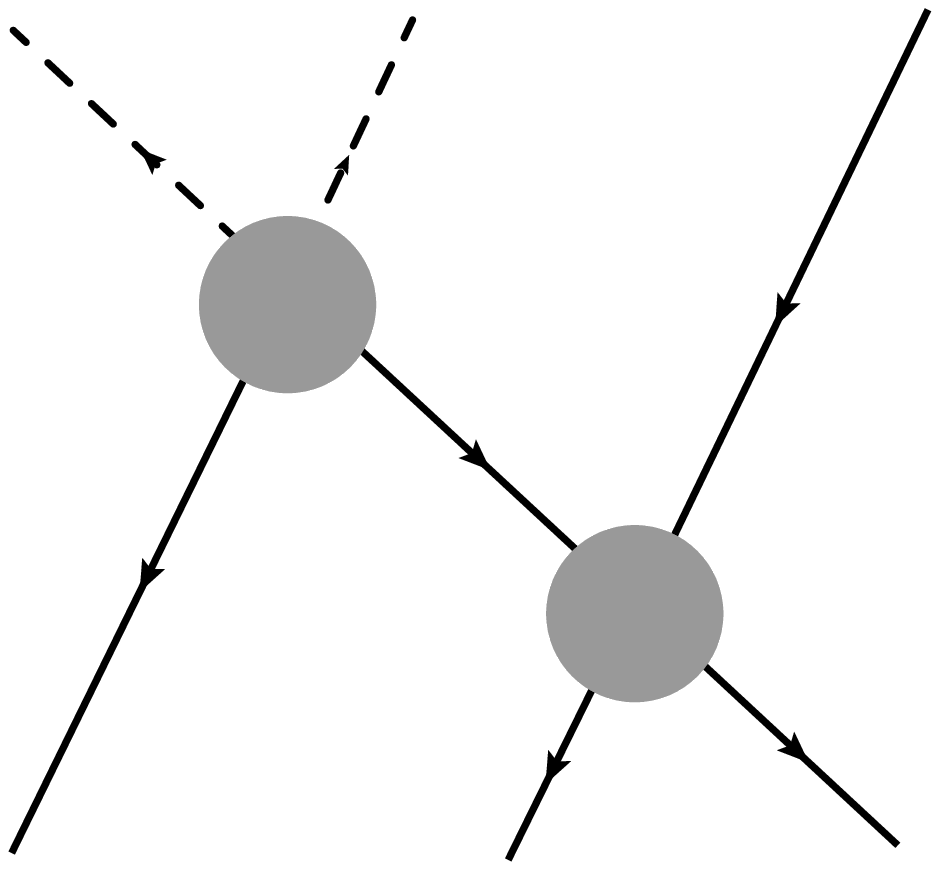}
\put(-82,63){\small$\bfrac{ren.}{tree}$}
 \put(-42,27){\footnotesize$new$}
 \put(-237,63){\footnotesize$new$}
 \put(-197,27){\footnotesize$new$}
\put(-393,63){\footnotesize$new$}
 \put(-353,27){\footnotesize$new$}
 \caption{\it 
 In this figure possible  connected  graphs   $\delta G_{c \,n}(\mathcal{W})$
that are made out of two $\delta \Gamma_{ren.tree}$ and $\Gamma_{new}$ vertices
are depicted. Only the first one on the left can be Wick-Contracted to the the operators of the sector, and it leads to a finite contribution when inserted in $\langle \bar{\mathcal{O}} \mathcal{O} \rangle$ due to the non-renormalization theorem of \cite{Fiamberti:2008sh,Sieg:2010tz}.}
 \label{connected}
\end{figure}

 \medskip

\begin{figure}[h!!]
   \centering
   \includegraphics[scale=0.5]{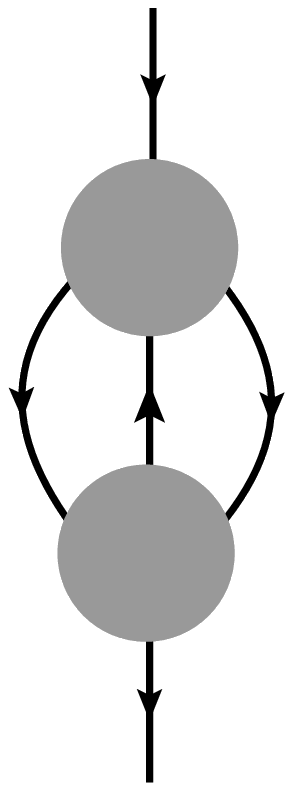}
   \qquad    \qquad    \qquad
        \includegraphics[scale=0.5]{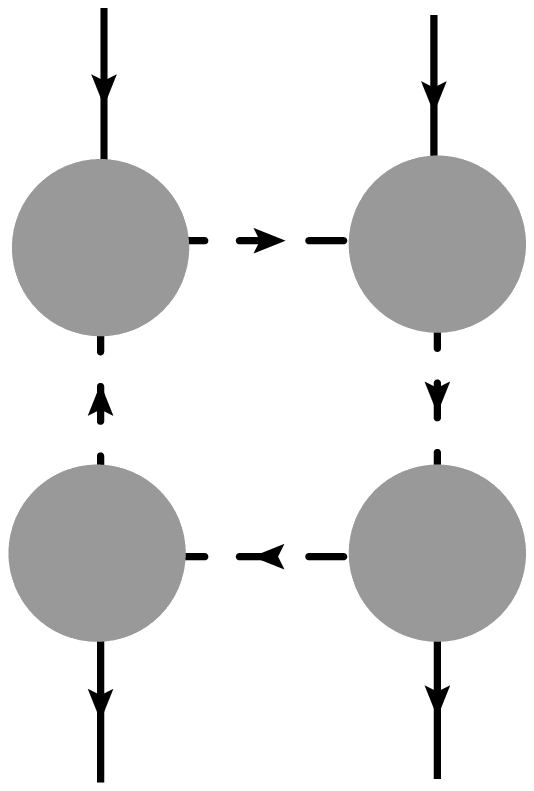}
   \qquad    \qquad    \qquad
        \includegraphics[scale=0.65]{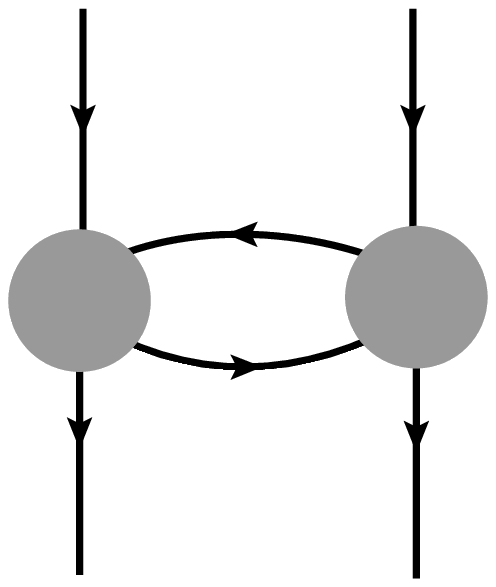}
  \caption{\it These are examples of diagrams that, although we make them from 1PI vertices, they are still 1PI as they  cannot be disjoint by cutting a single line. This means that we have already consider them in the previous section. }
 \label{1pi-already}
\end{figure}
 
 Examples of diagrams that correspond to connected  graphs
  $\delta G_{c \,n}(\mathcal{W})$ are depicted in Figure \ref{connected}.
  Examples of diagrams that correspond to
 1PI $\delta \Gamma_{n}(\mathcal{W})$  are depicted in Figure \ref{1pi-already}.
 The diagrams in Figure  \ref{1pi-already} should {\it not} be considered as arising from the contraction of two or more vertices (because this makes things difficult), but as coming from an insertion of a single vertex  that is created at some higher loop order.
Note that we have switched formalism from $\mathcal{N}=1$ superspace to the real $\mathbb{R}^{4|8}$ $\mathcal{N}=2$ superspace introduced in section \ref{superspace}. In Figures  \ref{connected} and  \ref{1pi-already}  the solid lines, now, depict the $\mathcal{N}=2$ chiral superfield strength $\mathcal{W}$ that
includes all the component fields in the $\mathcal{N}=2$ vector multiplet and  the dashed lines for the $\mathcal{N}=2$ fundamental hypermultiplet.

 \medskip

There are two classes of  $\delta \Gamma^{new}_{n}$, that we should discuss why they cannot  make $\delta G^{new}_{n} (\mathcal{W})$ that can contribute to the Hamiltonian
 when combined with themselves ($\delta \Gamma^{new}_{n}$) or $\delta \Gamma^{ren.tree}_{n}$. 
\begin{enumerate}
\item $\delta \Gamma^{new}_{n}(\mathcal{W}, Q)$: new vertices that include hypermultiplets $(Q,\tilde{Q})$ as external fields
\item $\delta \Gamma^{new}_{n}(\mathcal{W})$: new vertices with only $\mathcal{W}$'s (and $\bar{\mathcal{W}}$'s) as external fields
\end{enumerate}

 For the first class of  new vertices in the effective action that include $Q$'s the argument goes as follows.
In order to
hide all the $Q$'s inside loops (if not we cannot Wick-contract them to the sector), so that the  $\delta G_{c \,n}(\mathcal{W})$  has only $\mathcal{W}$ and  $\bar{\mathcal{W}}$ external fields and can be Wick-contracted to $\mathcal{O}$ and $\bar{\mathcal{O}}$,
we need to make a $Q$-loop! Such a diagram is always 1PI because there are always two internal $Q$ propagators. A  $Q$-loop cannot be disjoined by cutting a single line. 

\medskip

For the second class of  new vertices
 in the effective action, after careful inspection of all the possible {\it new effective vertices} available \eqref{effective}, we observe that:

%

\begin{itemize}
\item
Multiple vertices of the form $\Gamma_{new}^{can}$  \eqref{new-can-N2supersp}  also lead to finite contributions due to the non renormalization theorem, as explained in Section \ref{non-renorm-theorem}. See Figure \ref{connected6} and the discussion in Appendices \ref{6-loop} and \ref{non-ren-ex} for an example.
\item
Vertices of the form $\Gamma_{new}^{can}$ \eqref{new-can-N2supersp}  combined with $\delta \Gamma_{ren.tree}$ will also give  finite contributions when inserted in the operator renormalization diagram due to the non-renormalization theorem of \cite{Fiamberti:2008sh,Sieg:2010tz}.
Note that for the $SU(2,1|2)$ sector Hamiltonian there are {\it no vertices coming from the superpotential} that can be attached to $\Gamma_{new}^{can}$ externally in such a way that it can be afterwards Wick-contracted to the operator. Such an observation was already used in Section \ref{1-loop}. One more concrete example is given in
Appendix
\ref{Four-loops}.

\item
For combinations of multiple vertices from \eqref{effective} with $\Psi$'s \eqref{psi's} there are two possibilities.
They can either not be Wick-contracted to $\langle \bar{\mathcal{O}} \mathcal{O} \rangle$ in the sector at all,
 or if all the $\Psi$'s are internally contracted can only make 1PI  vertices of the form $\Gamma_{new}^{can}$  \eqref{new-can-N2supersp}, that will not contribute due to the non-renormalization theorem of \cite{Fiamberti:2008sh,Sieg:2010tz}.
\end{itemize}
For some concrete examples at four-, five- and six-loops see Appendix \ref{multi-app}.

\bigskip

\subsection{The derivatives ``commute'' with the dressing of the skeletons}
\label{derivatives-all-loop}

To complete our argument we need to explain why all the observations we made in the previous sections still hold  even when the operator $\mathcal{O}$ includes derivatives.
The only difference in the calculation between operator renormalization with and without derivatives is that the final integral one has to perform has extra momenta in the numerator in the case of an operator with derivatives. For each derivative in the operator, we have one momentum in the numerator of the final integrand. We argue in this section that operators with derivatives in the $SU(2,1|2)$ sector create in the numerators of the loop integrals
{\it traceless symmetric products of momenta}  which {\it do not alter the degree of divergence of the loop integrals}.

\bigskip

The proof of this statement goes as follows.
For an operator $\mathcal{O}_L = \tr\left(\mathcal{W}^L \right)$ in $SU(2,1|2)$ that does not include any derivatives the integrals that appear when we try to compute the two-point function $\langle \bar{\mathcal{O}}  \mathcal{O} \rangle$ will have the form
\be
 \int dq_1\cdots dq_k  \, \mathcal{I}\left(q_1^2,\dots,q_k^2\right)
\ee
where $\mathcal{I}\left(\{q^2_i \}\right)$ is a scalar under Lorentz transformations.
The integral, when considering the renormalization of an operator composed of the same fields as before but with extra derivatives, schematically $\mathcal{O}^n_L = \tr\left(\mathcal{D}_{+ \dot{\alpha}}^n\mathcal{W}^L \right)$\footnote{Of course it is important where the derivatives are, $\mathcal{O}^{\{n\}}_L = \tr\left(\mathcal{D}^{n_1}\mathcal{W} \, \mathcal{D}^{n_2}\mathcal{W} \cdots\right)$,  in which site of the spin chain with $\mathcal{O}_L = \tr\left(\mathcal{W}^L \right)$ vacuum.},
 will be obtained by inserting momenta (one for each derivative) and will have the form 
\be
 \int dq_1\cdots dq_\ell  \, \mathcal{I}\left(q_1^2,\dots,q_\ell^2\right)q_1^{+ \dot\alpha_1}
 q_2^{+ \dot\alpha_2} \cdots 
 \, \, .
\ee
{\it Lorentz symmetry} does not allow for an integral that is not a scalar to give a non zero answer.
Performing the integrals with the extra momenta in the numerators can either give zero or after partial integration an integral proportional to
\be
\left( q_{ext_1}^{+ \dot\alpha} q_{ext_2}^{+ \dot\beta}  \cdots  \right)\int dq_1\cdots dq_\ell  \, \mathcal{I}\left(q_1^2,\dots,q_\ell^2\right) \, ,
\ee
where the momenta $q_{ext_1}^{+ \dot\alpha}$, $q_{ext_2}^{+ \dot\beta}$, \dots are external to the loops we are integrating over.
This means that due to {\it Lorentz invariance} ({\it covariance})  {\it the $SU(2,1|2)$ sector derivatives have to go out} of the integral!
The integral we have to evaluate will have the same divergence structure as the integral without derivatives.
Derivatives in the $SU(2,1|2)$ sector 
create 
{\it traceless symmetric products of momenta}  which {\it do not alter the degree of divergence of the loop integrals}.

\bigskip

Applying what we just learned the first thing to notice is that
the new effective vertices of  the form $\tr \left( \mathcal{W}^n \bar{\mathcal{W}}^n\right)$ will still lead to finite integrals and will not contribute to the Hamiltonian.
We have already seen an example of this  in the three-loop section \ref{Three-loops} and here we generalize this statement for any vertex of the form \eqref{new-can-N2supersp}.

\medskip

Thus, even when we take into account operators with derivatives, only the skeleton diagrams can be inserted in the two-point functions $\langle \bar{\mathcal{O}}  \mathcal{O} \rangle$ and lead to logarithmic divergencies.
In fact we should be careful and note that the argument that we are giving here is only for plain derivatives $\partial_{+\dot+}$; not for covariant derivatives. The generalization to include gauge boson emission processes is incorporated by the use of the background field formalism that guarantees gauge invariance.

\bigskip

Given the fact that only skeleton diagrams can be inserted in the two-point functions $\langle \bar{\mathcal{O}}  \mathcal{O} \rangle$
 what we do is to begin with the tree level integral of
\be
\label{tree-squem}
\langle \bar{\mathcal{O}} | G_{c \,n}^{(tree)} (\mathcal{W})|  \mathcal{O} \rangle \, 
\ee
and at $\ell$ loops replace it with
\be
\label{ell-squem}
\langle \bar{\mathcal{O}} |\delta G_{c \,n}^{(\ell)} (\mathcal{W})|  \mathcal{O}  \rangle
\ee
 where $\delta G_{c \,n}^{(\ell)}(\mathcal{W})$ are the connected  off-shell $n$-point functions at $\ell$ loops with external $\mathcal{W}$'s and $\bar{\mathcal{W}}$'s, defined in \eqref{GcV}.
 In the previous sections we have shown that \eqref{all-nice}
\be
\label{Gloop-Gtree}
\delta  G_{c \,n}^{(\ell)}(\mathcal{W};g) =G_{c \,n}^{(tree)} (\mathcal{W};\newg)
\ee
This means that the leading divergent part of the integrals, that appear in the $\ell$-loop Hamiltonian calculations for some particular distribution external number of derivatives (momenta) is identical, 
{\it as a function of momenta, to  the ``tree level''}  integrals, {\it i.e.} identical to the ones in $\mathcal{N}=4$ the first time they appear.  
The main lesson of this section could be phrased as the statement that {\it the derivatives ``commute'' with the operation of dressing the skeleton diagrams}, and this concludes our argument for $\mathcal{N}=2$ superconformal gauge theories.

\bigskip

\section{Conclusions and discussion}

In this paper, building up on the work of
\cite{Gadde:2010zi,Pomoni:2011jj,Gadde:2010ku,Liendo:2011xb,Gadde:2012rv},
we have discussed first why any ${\cal N}=2$ superconformal gauge theory (including the
${\cal N}=4$ SYM) contains an $SU(2,1|2)$ sector that is made out of {\it only fields in the vector multiplet} and  that is closed to all loops under renormalization.
This statement is valid to all orders of the `t Hooft coupling constant in the planar limit, and since the `t Hooft coupling
expansion is believed to converge \cite{'tHooft:1982tz}, it is also a true statement at finite `t Hooft coupling.
We have then presented a diagrammatic argument  that the asymptotic $SU(2,1|2)$ Hamiltonian of any ${\cal N}=2$ superconformal gauge theory is identical at all loops to that of ${\cal N}=4$ SYM
\be
\nonumber
H_{\mathcal{N}=2}(g) = H_{\mathcal{N}=4}(\newg) \qquad \mbox{with} \qquad \newg = \sqrt{f(g^2)} \, ,
\ee
 up to a redefinition of the coupling constant $g^2 \rightarrow f(g^2)= g^2 + \mathcal{O}\left(g^6\right)$. 
We wish to insist on a {\it disclaimer}: the Hamiltonian that we have been discussing here is the {\it asymptotic} Hamiltonian! It {\it does not include wrapping corrections} \cite{Sieg:2005kd} and it can only compute the anomalous dimensions of sufficiently long operators. It can compute the anomalous dimensions of operators that correspond to spin chain states with their number of sites $L \geq \ell+1$ being bigger than the range of the interaction which is specified by the number of loops $\ell$.
We leave the study of  wrapping corrections for future work.

\bigskip

To finish the job and actually be able to compute the spectrum of ${\cal N}=2$ superconformal gauge theories we need to calculate the function $f(g^2)$ (or $\delta Z(g)$). One way to obtain $\delta Z(g)$ is to perform Feynman diagram computations and compute the difference in the self-energy  of  $\mathcal{W}$ in ${\cal N}=4$ and ${\cal N}=2$ superconformal gauge theories.
In fact in \cite{Pomoni:2011jj} one can already find the answer for the 2-loop self-energy (3-loop Hamiltonian).
Of course at some point this method will run out of steam as Feynman diagram computations will get very hard quite fast.
Alternatively, one can consider  the circular Wilson loop\footnote{In \cite{Andree:2010na} the calculation of the  circular Wilson was loop  performed up to three loops but in components. For this program to have any hope of success one will have to proceed using $\mathcal{N}=2$ superspace.}, for which exact results can be obtained using localization  \cite{Pestun:2007rz}. 
When the circular Wilson loop is calculated diagrammatically the final result will deviate from the $\mathcal{N}=4$ one solely due to the universal function of the coupling $\delta Z(g)$ (or  $f(g^2)$).
For example, one should be able to extract  $\delta Z(g)$ for the $\mathcal{N}=2$ SCQCD from the result of 
 \cite{Passerini:2011fe} and for the $\mathbb{Z}_2$ interpolating quiver from the result of \cite{ Wenbin}\footnote{This was done after the first version of this paper appeared in the arXiv in \cite{Mitev:2014yba}.}.
Finally, we should also be able to compare with the cusp anomalous dimensions
\cite{Belitsky:2003ys} where the function $f(g^2)$ should also appear as a universal function. 

\bigskip

Our result implies that  any planar $\mathcal{N}=2$ superconformal gauge theory in the $SU(2,1|2)$ sector is integrable, with its integrability inherited directly from planar $\mathcal{N}=4$ SYM.
Given this result, we should also address the question of which particular properties make a gauge theory integrable.
We were able to formulate our argument by comparing the planar $\mathcal{N}=2$ Hamiltonian with the ${\cal N}=4$ SYM one, and thus {\it planarity} is essential and irreplaceable. Moreover, for our argument {\it the choice of the sector} was crucial, and in particular the fact that all the fields that compose the operators in the $SU(2,1|2)$ sector are in the $\mathcal{N}=2$ {\it vector multiplet}! In order, though, to be able to restrict to a sector with only fields in the vector multiplet, we had to restrict Lorentz indices to $\alpha = +$, highest weight states (symmetric representations of $SU(2)_{\alpha}$). This restriction ``protected'' the operators from possible corrections coming from new effective vertices. Only terms that exist in the action already at tree level (of course renormalized) were allowed to contribute.
Finally, {\it gauge invariance} (and renormalizability) played a critical role. Everything was renormalized with a single $Z(g)$ function, when BFM was employed.

\bigskip

A key element for the integrability of the $SU(2,1|2)$ sector was the fact that
all the fields composing the sector are in the $\mathcal{N}=2$ {\it vector multiplet}. It is thus very compelling to
look for possible integrable  subsectors in other gauge theories with the same property. 
 For  $\mathcal{N}=1$ superconformal gauge theories the $\mathcal{N}=1$ {\it vector multiplet} contains the gluon and the gluino, and the
biggest subsector with fields only in the vector multiplet is an
 $SU(2,1|1)$  sector that is closed to all loops and contains
 \be
 \label{sector-fieldsN1}
\lambda_{+}      \, ,    \quad   \mathcal{F}_{++}  
  \, ,    \quad 
 \mathcal{D}_{+ \dot\alpha}  \, ,
\ee
with
\be
\label{sectorcongN1}
\Delta = 2 j - r    \qquad \forall \,  \mathcal{O} \in SU(2,1|1) \, .
\ee
Similarly, for $\mathcal{N}=0$  superconformal gauge theories we should consider the
 $SU(2,1)$  sector that is closed to all loops and contains 
 \be
 \label{sector-fieldsN0}
   \mathcal{F}_{++}  
  \, ,    \quad 
 \mathcal{D}_{+ \dot\alpha}  \, ,
\ee
with
\be
\label{sectorcongN0}
\Delta = 2 j    \qquad \forall \,  \mathcal{O} \in SU(2,1) \, .
\ee
As we stressed in section \ref{sector}, there is no reason why   \eqref{sectorcongN1} and \eqref{sectorcongN0} should persist in perturbation theory. In fact they will be violated by corrections of the order of $g^2$. However, for the fields outside of the sectors \eqref{sector-fieldsN1} and \eqref{sector-fieldsN0} the equalities \eqref{sectorcongN1} and \eqref{sectorcongN0} are violated already classically by an integer, thus perturbative corrections will never be big enough to allow them to enter the sector.

\medskip

Gauge invariance (when the BFM is used), and supersymmetry (for the ${\cal N}=1$ case), imply that in these sectors all the fields that compose the operators are renormalized with a single $Z(g)$.
Up to three-loops we can see that the statement \eqref{statement} goes through also for any $\mathcal{N}=1$ and any $\mathcal{N}=0$  superconformal gauge theory.  This is work in progress. 
 To cut a long story short, if we could succeed in showing that the new effective vertices that appear in the effective actions cannot contribute to the
$SU(2,1|1)$ sector of ${\cal N}=1$ and to the $SU(2,1)$ sector of ${\cal N}=0$ gauge theories, we will have shown that \eqref{statement} generalizes for any superconformal gauge theory.  Up to tree loops this is very simple.
 The missing element in pushing \eqref{statement} to higher loops is that
we do not have a complete classification of all the possible new vertices that can in principle appear in the effective action.

\bigskip

We would like to conclude our paper with a comment on $\mathcal{N}=4$ SYM.
It is believed, due to different types of calculations, that  the magnon dispersion relation for  $\mathcal{N}=4$ (see \cite{Sieg:2010jt,Sieg:2010tz} and references therein)
\be
E(p;g) =  \sqrt{1+8 h(g) \sin^{2}\left(\frac{p}{2}\right)  }
\ee
gets no corrections in perturbation theory and $h(g)=g^2$. This result is definitely tied to the fact that if one uses the appropriate language (that might be the light-cone superspace formalism of \cite{Brink:1982pd,Mandelstam:1982cb}\footnote{See \cite{Ananth:2012tf} for a modern presentation that is actually devoted on correlation functions of composite gauge-invariant operators of  $\mathcal{N}=4$  SYM.}) there are no corrections for $g$ at all, {\it i.e.} $Z(g)=1$ for $\mathcal{N}=4$ SYM. This would mean that the computation of the Hamiltonian would  contain only bare skeleton diagrams - and combinatorial factors.
According to the way of thinking that we have presented here, one should start from the effective action, combined with the fact that  $Z(g)=1$ and then try to obtain the Hamiltonian. This strategy should make it possible to calculate the Hamiltonian of $\mathcal{N}=4$ to more (maybe all) loops, at least in some subsectors.

\bigskip

\section*{Acknowledgments}
We are grateful to   Leonardo Rastelli and Christoph Sieg for their important impact into our work, their kind support and mentoring.
It is also a great pleasure to thank Isabella Bierenbaum, Nadav Drukker, Burkhard Eden, Valentina Forini,  Manuela Kulaxizi,  Pedro Liendo, Carlo Meneghelli, Vladimir Mitev,   Martin Rocek and 
Katy Tschann-Grimm
for useful discussions and correspondence.
This work is part of a long term project that has been partially supported by the Humboldt Foundation, the Marie Curie grant FP7-PEOPLE-2010-RG, DESY and the Initial Training Network GATIS.

\bigskip

\bigskip

\appendix

\section{Multi-vertex examples from higher loops}
\label{multi-app}

In this Appendix we present a few examples of multi-vertex insertions
with one or more $\delta  \Gamma_{new}$ vertices,
and explain why they cannot contribute to anomalous dimensions.
Up to three loops only single $\delta  \Gamma_{new}$ vertex diagrams appear and we have presented the reasons why they can not contribute to the Hamiltonian in Section \ref{123}. But, from four loops and on, combinations of such single vertices (with $\delta \Gamma_{ren. \, tree}$ or $\delta  \Gamma_{new}$) have to be considered. This was addressed in Section \ref{1PI2connected}, but we think that it is very useful to supplement the arguments there by some explicit examples. This is what we do in this Appendix.

\subsection{Four-loops}
\label{Four-loops}
\begin{figure}[h!!]
   \centering 
     \includegraphics[scale=0.5]{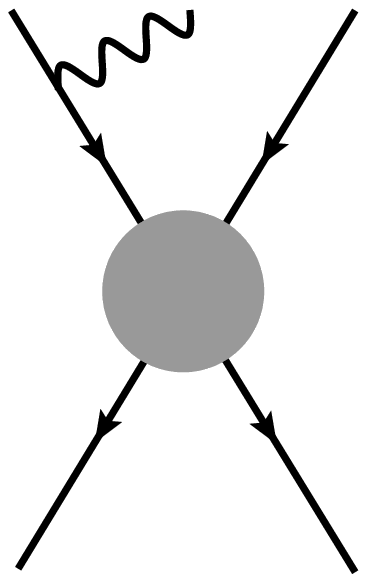}
     \qquad \qquad 
        \includegraphics[scale=0.5]{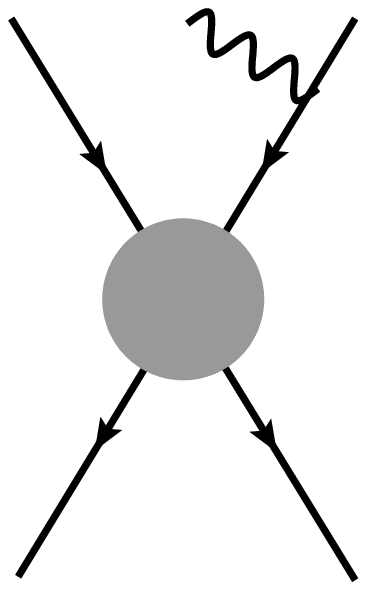}
      \put(-35,39){\tiny2-$loop$}
            \put(-76,37){\large {\bf +}}
                \put(-134.5,39){\tiny2-$loop$}
                           \put(26,37){\large {\bf $=\quad0$}}
 \caption{\it Examples of diagrams that appear at order $g^7$ (three loops and a half). They are made out of a $\delta\Gamma_{new}^{(2)}$ and a tree level cubic vertex. These diagrams will add up to zero in $\mathcal{N}=1$ superspace. They are examples of the ``{\it miraculous cancellations}'' that have to happen in $\mathcal{N}=1$ superspace because we are not keeping the whole symmetry manifest.}
 \label{connected41}
\end{figure}

Some of the first diagrams that may come to mind, for someone who is used to $\mathcal{N}=1$ superspace, are the ones depicted in Figure \ref{connected41}. These diagrams appear at order $g^7$ (three loops and a half) and are made out of a $\delta\Gamma_{new}^{(2)}$ and a tree level cubic vertex. 
These diagrams from the $\mathcal{N}=1$ superspace point of view look like they may contribute to logarithmic divergences. But, after summing them all up, the $\mathcal{N}=1$ superspace practitioner will discover that they all add up to zero. This is because the two different vertices $\tr \left( \bar{\Phi}V\Phi \right)$ and $\tr \left( \bar{\Phi}\Phi V \right)$  differ by a minus sign. The reader can find many such examples explicitly worked out in \cite{Sieg:2010tz}.

\subsection{Five-loops}

\begin{figure}[h!!]
   \centering 
     \includegraphics[scale=0.45]{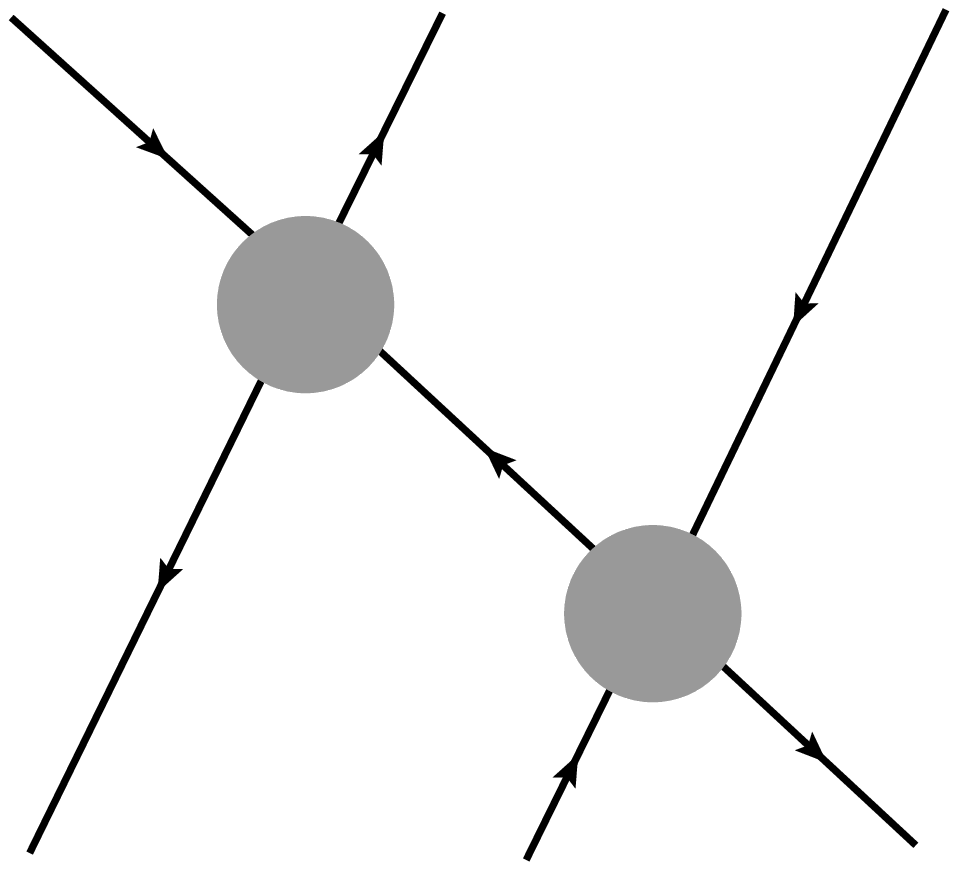}
 \put(-48.5,31){\tiny1-$loop$}
 \put(-94,71){\tiny2-$loop$}
 \caption{\it This connected diagram is made out of a one-loop new 1PI vertex and a two-loop new 1PI vertex.
 This diagram cannot be planarly Wick-contracted.}
 \label{connected5}
\end{figure}

At five-loops and order $g^{10}$ another new type of diagram that we wish to examine appears, depicted in Figure \ref{connected5}. This is a  connected diagram that is not 1PI because it is made by gluing two vertices with a single propagator.
This diagram will not contribute for many reasons. One reason is that it  cannot be planarly Wick-contracted to the operators in the $SU(2,1|2)$ sector.

\medskip

One might have similar worries for the case of combining by gluing a single line between 
new effective vertices of the form $W^{\alpha}W_\alpha\bar{W}^{\dot{\alpha}}\bar W_{\dot{\alpha}}$. One could try to generalize the Lorentz invariance of Section \ref{Three-loops}, but this is not a good strategy. This is the moment when one should just abandon  $\mathcal{N}=1$ superspace and realize that these vertices are just inside the $\mathcal{N}=2$  effective action in equation  \eqref{effective}.

\subsection{Six-loops}
\label{6-loop}

\begin{figure}[h!!]
   \centering 
     \includegraphics[scale=0.45]{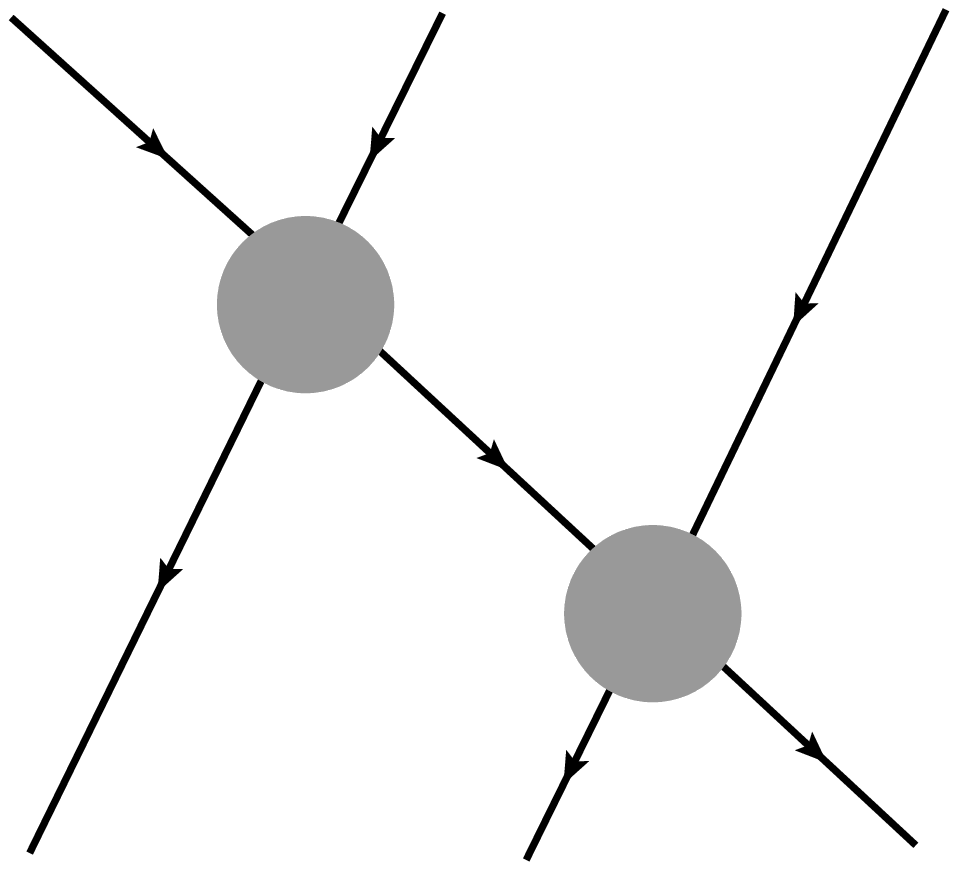}
 \put(-48.5,31){\tiny2-$loop$}
 \put(-94,71){\tiny2-$loop$}
 \caption{\it This connected diagram cannot contribute to the divergencies due to the non-renormalization theorem of \cite{Fiamberti:2008sh,Sieg:2010tz}.}
 \label{connected6}
\end{figure}

In Figure \ref{connected6} we give one last example of a $g^{12}$ diagram (six-loop Hamiltonian). This diagram is also new in the sense that at six loops it is the first time we can combine two $\delta\Gamma_{new}^{(2)}$ vertices to make a connected graph. This connected diagram cannot contribute to the divergences due to the non-renormalization theorem of \cite{Fiamberti:2008sh,Sieg:2010tz}. The explicit powercounting is performed in the next section of the Appendix \ref{non-ren-ex} (Figure  \ref{powercounting} ($b$)).

\section{Explicit examples of powercounting}
\label{non-ren-ex}
In this section of the Appendix we perform the  powercounting explicitly for two particular examples in order to demonstrate how/why the non-renormalization theorem of \cite{Fiamberti:2008sh,Sieg:2010tz} works. The non-expert reader might have to first read \cite{Gates:1983nr,Fiamberti:2008sh,Sieg:2010tz}.

 \begin{figure}[h!!]
   \centering 
     \includegraphics[scale=0.55]{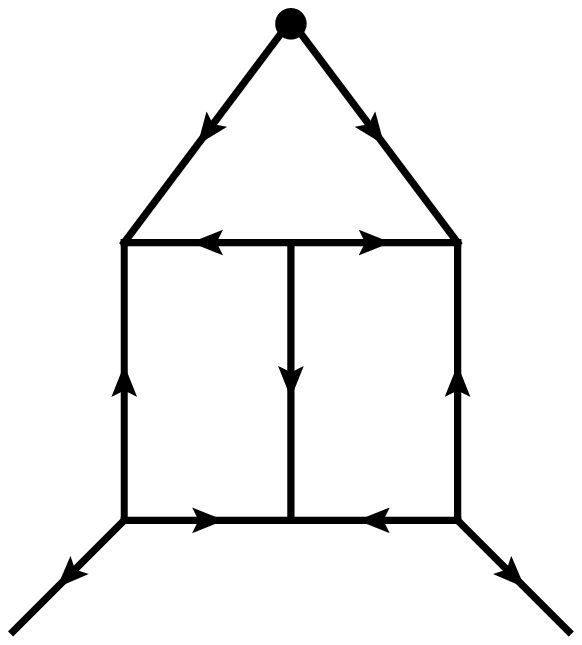}
          \put(-44,66){\footnotesize$\bar{D}^2$}
 \put(-56,66){\footnotesize$\bar{D}^2$}
      \put(-44,10){\footnotesize${D}^2$}
 \put(-56,10){\footnotesize${D}^2$}
 \put(-78,67){\footnotesize$D^2$}
  \put(-21,67){\footnotesize$D^2$}
   \put(-84,54){\footnotesize${D}^2$}
  \put(-18,54){\footnotesize${D}^2$}
  \put(-40,93){\footnotesize$\bar{D}^2$}
     \put(-84,20){\footnotesize$\bar{D}^2$}
  \put(-18,20){\footnotesize$\bar{D}^2$}
  \put(-50,-10){\footnotesize($a$)}  
  \qquad    \qquad
  \qquad   
    \includegraphics[scale=0.55]{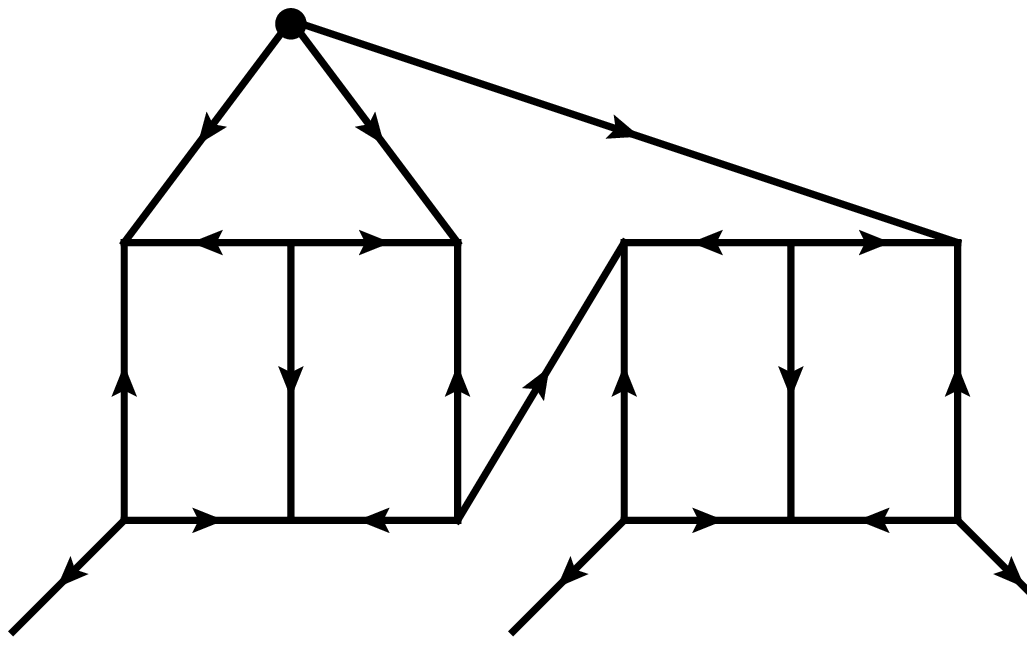}
      \put(-100,-10){\footnotesize($b$)}  
 \put(-44,54){\footnotesize$\bar{D}^2$}
 \put(-57,54){\footnotesize$\bar{D}^2$}
      \put(-44,10){\footnotesize${D}^2$}
 \put(-56,10){\footnotesize${D}^2$}
 \put(-84,60){\footnotesize$D^2$}
  \put(-23,67){\footnotesize$D^2$}
   \put(-72,54){\footnotesize${D}^2$}
  \put(-18,54){\footnotesize${D}^2$}
     \put(-84,20){\footnotesize$\bar{D}^2$}
  \put(-18,20){\footnotesize$\bar{D}^2$}
  \put(-140,93){\footnotesize$\bar{D}^2$}
    \put(-115,96){\footnotesize$\bar{D}^2$}
    \put(-124,66){\footnotesize$\bar{D}^2$}
 \put(-136,66){\footnotesize$\bar{D}^2$}
      \put(-124,10){\footnotesize${D}^2$}
 \put(-136,10){\footnotesize${D}^2$}
 \put(-158,67){\footnotesize$D^2$}
  \put(-101,67){\footnotesize$D^2$}
   \put(-164,54){\footnotesize${D}^2$}
  \put(-98,54){\footnotesize${D}^2$}
     \put(-164,20){\footnotesize$\bar{D}^2$}
  \put(-96,20){\footnotesize$\bar{D}^2$}
    \put(-110,21){\footnotesize$\bar{D}^2$}
 \caption{\it In this Figure we show two examples of powercounting. The solid lines depict chiral superfields and the black dot the operator insertion. Only the ``effective number'' of $D$'s (= number of $D$'s - twice the number of scalar propagators not belonging to any loop) \cite{Fiamberti:2008sh} is put in the Figure.}
 \label{powercounting}
\end{figure}

First we wish to show why the diagram depicted in Figure \ref{powercounting} ($a$) will not contribute to anomalous dimensions.
Note that, as we discussed in Section \ref{non-renorm-theorem}  there is one $\bar{D}^2$ missing from the operator and also two $\bar{D}^2$ that could go to the external legs are not even shown in the figure because they are external to the loops.
As one can see from the figure we have 5 $\bar{D}^2$'s and  6 ${D}^2$'s for which we have to perform the $D$-algebra.
The diagram has 3 loops, which means that there are 3 integrals $\int d^4\theta$ that have to be performed and these integrals will eat  up 3 $\bar{D}^2$'s and  3 ${D}^2$'s.
After the integrations we will be left with 2 $\bar{D}^2$'s and  3 ${D}^2$'s that can create a maximum of 2 $p^2$ in the numerator.
The diagram moreover contains
9 propagators and the
 3 loops will lead to 3 momentum integrals.  All in all, 
 \be
 \int^{\Lambda} \left[d^4p \right]^3 \frac{\left(p^2\right)^2}{\left( p^2 \right)^9} \sim \frac{1}{\Lambda^2}
 \ee
 the superficial degree of divergence is $-2$,
which means that the diagram is not divergent.

\medskip

We then consider the diagram depicted in Figure \ref{powercounting} ($b$).
As before  one $\bar{D}^2$ is missing from the operator and three $\bar{D}^2$ that could go to the external legs are not even shown in the figure.
It contains 11 $\bar{D}^2$'s and  12 ${D}^2$'s
 which the 6 loop integrals $\int d^4\theta$  will reduce to  5 $\bar{D}^2$'s and  6 ${D}^2$'s. These $D$'s can only create 5 $p^2$ in the numerator.
The diagram contains
18 propagators and 
 6 loops.
Finally, we find that the superficial  degree of divergence
 \be
\int^{\Lambda}   \left[d^4p \right]^6 \frac{\left(p^2\right)^5}{\left( p^2 \right)^{18}} \sim \frac{1}{\Lambda^2}
 \ee
is $-2$ which again means that the diagram is not divergent.

%


\bigskip


\bigskip



\begin{thebibliography}{99}


\addtolength{\rightmargin}{0.1in} 
\addtolength{\leftmargin}{0.1in} 
\setlength{\itemindent}{-0.1in}


\bibitem{Beisert:2010jr} 
  N.~Beisert, C.~Ahn, L.~F.~Alday, Z.~Bajnok, J.~M.~Drummond, L.~Freyhult, N.~Gromov and R.~A.~Janik {\it et al.},
  ``Review of AdS/CFT Integrability: An Overview,''
  Lett.\ Math.\ Phys.\  {\bf 99}, 3 (2012)
  [arXiv:1012.3982 [hep-th]].



\bibitem{Pestun:2007rz} 
  V.~Pestun,
  ``Localization of gauge theory on a four-sphere and supersymmetric Wilson loops,''
  Commun.\ Math.\ Phys.\  {\bf 313}, 71 (2012)
  [arXiv:0712.2824 [hep-th]].

\bibitem{Grana:2001xn} 
  M.~Grana and J.~Polchinski,
  ``Gauge / gravity duals with holomorphic dilaton,''
  Phys.\ Rev.\ D {\bf 65}, 126005 (2002)
  [hep-th/0106014].


\bibitem{Lin:2004nb} 
  H.~Lin, O.~Lunin and J.~M.~Maldacena,
  ``Bubbling AdS space and 1/2 BPS geometries,''
  JHEP {\bf 0410}, 025 (2004)
  [hep-th/0409174].

\bibitem{Gaiotto:2009gz} 
  D.~Gaiotto and J.~Maldacena,
  ``The Gravity duals of N=2 superconformal field theories,''
  JHEP {\bf 1210}, 189 (2012)
  [arXiv:0904.4466 [hep-th]].


\bibitem{Gadde:2009dj} 
  A.~Gadde, E.~Pomoni and L.~Rastelli,
  ``The Veneziano Limit of N = 2 Superconformal QCD: Towards the String Dual of N = 2 SU(N(c)) SYM with N(f) = 2 N(c),''
  arXiv:0912.4918 [hep-th].
  
\bibitem{Colgain:2011hb} 
  E.~O Colgain and B.~Stefanski, Jr.,
  ``A search for AdS5 X S2 IIB supergravity solutions dual to N = 2 SCFTs,''
  JHEP {\bf 1110}, 061 (2011)
  [arXiv:1107.5763 [hep-th]].

  
  
\bibitem{Aharony:2012tz} 
  O.~Aharony, L.~Berdichevsky and M.~Berkooz,
  ``4d N=2 superconformal linear quivers with type IIA duals,''
  JHEP {\bf 1208}, 131 (2012)
  [arXiv:1206.5916 [hep-th]].
  
\bibitem{Stefanski:2013osa} 
  B.~Stefa?ski,
  ``Supermembrane actions for Gaiotto-Maldacena backgrounds,''
  Nucl.\ Phys.\ B {\bf 883}, 581 (2014)
  [arXiv:1308.2789 [hep-th]].

  

\bibitem{Gadde:2010zi} 
  A.~Gadde, E.~Pomoni and L.~Rastelli,
  ``Spin Chains in N=2 Superconformal Theories: From the $Z_2$ Quiver to Superconformal QCD,''
  JHEP {\bf 1206}, 107 (2012)
  [arXiv:1006.0015 [hep-th]].
  
  
\bibitem{Pomoni:2011jj} 
  E.~Pomoni and C.~Sieg,
  ``From N=4 gauge theory to N=2 conformal QCD: three-loop mixing of scalar composite operators,''
  arXiv:1105.3487 [hep-th].
  
\bibitem{Gadde:2010ku} 
  A.~Gadde and L.~Rastelli,
  ``Twisted Magnons,''
  JHEP {\bf 1204}, 053 (2012)
  [arXiv:1012.2097 [hep-th]].

  
\bibitem{Liendo:2011xb} 
  P.~Liendo, E.~Pomoni and L.~Rastelli,
  ``The Complete One-Loop Dilation Operator of N=2 SuperConformal QCD,''
  JHEP {\bf 1207}, 003 (2012)
  [arXiv:1105.3972 [hep-th]].
  
\bibitem{Gadde:2012rv} 
  A.~Gadde, P.~Liendo, L.~Rastelli and W.~Yan,
  ``On the Integrability of Planar $N=2$ Superconformal Gauge Theories,''
  JHEP {\bf 1308}, 015 (2013)
  [arXiv:1211.0271 [hep-th]].

  
\bibitem{Korchemsky:2010kj} 
  G.~P.~Korchemsky,
  ``Review of AdS/CFT Integrability, Chapter IV.4: Integrability in QCD and N<4 SYM,''
  Lett.\ Math.\ Phys.\  {\bf 99}, 425 (2012)
  [arXiv:1012.4000 [hep-th]].

\bibitem{Belitsky:2004sf} 
  A.~V.~Belitsky, G.~P.~Korchemsky and D.~Mueller,
  ``Integrability in Yang-Mills theory on the light cone beyond leading order,''
  Phys.\ Rev.\ Lett.\  {\bf 94}, 151603 (2005)
  [hep-th/0412054].
  
\bibitem{Belitsky:2005bu} 
  A.~V.~Belitsky, G.~P.~Korchemsky and D.~Mueller,
 ``Integrability of two-loop dilatation operator in gauge theories,''
  Nucl.\ Phys.\ B {\bf 735}, 17 (2006)
  [hep-th/0509121].
  
  
\bibitem{Beisert:2004fv} 
  N.~Beisert, G.~Ferretti, R.~Heise and K.~Zarembo,
  ``One-loop QCD spin chain and its spectrum,''
  Nucl.\ Phys.\ B {\bf 717}, 137 (2005)
  [hep-th/0412029].

\bibitem{Ferretti:2004ba} 
  G.~Ferretti, R.~Heise and K.~Zarembo,
  ``New integrable structures in large-N QCD,''
  Phys.\ Rev.\ D {\bf 70}, 074024 (2004)
  [hep-th/0404187].



  
\bibitem{Poland:2011kg} 
  D.~Poland and D.~Simmons-Duffin,
  ``N=1 SQCD and the Transverse Field Ising Model,''
  JHEP {\bf 1202}, 009 (2012)
  [arXiv:1104.1425 [hep-th]].


\bibitem{Liendo:2011wc} 
  P.~Liendo and L.~Rastelli,
  ``The Complete One-loop Spin Chain of N =1 SQCD,''
  JHEP {\bf 1210}, 117 (2012)
  [arXiv:1111.5290 [hep-th]].
  


\bibitem{Zoubos:2010kh} 
  K.~Zoubos,
  ``Review of AdS/CFT Integrability, Chapter IV.2: Deformations, Orbifolds and Open Boundaries,''
  Lett.\ Math.\ Phys.\  {\bf 99}, 375 (2012)
  [arXiv:1012.3998 [hep-th]].

\bibitem{Fiamberti:2008sh} 
  F.~Fiamberti, A.~Santambrogio, C.~Sieg and D.~Zanon,
  ``Anomalous dimension with wrapping at four loops in N=4 SYM,''
  Nucl.\ Phys.\ B {\bf 805}, 231 (2008)
  [arXiv:0806.2095 [hep-th]].


  

\bibitem{Sieg:2010tz} 
  C.~Sieg,
  ``Superspace computation of the three-loop dilatation operator of N=4 SYM theory,''
  Phys.\ Rev.\ D {\bf 84}, 045014 (2011)
  [arXiv:1008.3351 [hep-th]].
  
  
    
\bibitem{Buchbinder:1999jn} 
  I.~L.~Buchbinder, S.~M.~Kuzenko and A.~A.~Tseytlin,
  ``On low-energy effective actions in N=2, N=4 superconformal theories in four-dimensions,''
  Phys.\ Rev.\ D {\bf 62}, 045001 (2000)
  [hep-th/9911221].
  
\bibitem{Kuzenko:2007cg} 
  S.~M.~Kuzenko and S.~J.~Tyler,
  ``Supersymmetric Euler-Heisenberg effective action: Two-loop results,''
  JHEP {\bf 0705}, 081 (2007)
  [hep-th/0703269].


\bibitem{Kuzenko:2003eb} 
  S.~M.~Kuzenko and I.~N.~McArthur,
  ``On the background field method beyond one-loop: A Manifestly covariant derivative expansion in superYang-Mills theories,''
  JHEP {\bf 0305}, 015 (2003)
  [hep-th/0302205].

  
\bibitem{Minahan:2010js} 
  J.~A.~Minahan,
  ``Review of AdS/CFT Integrability, Chapter I.1: Spin Chains in N=4 Super Yang-Mills,''
  Lett.\ Math.\ Phys.\  {\bf 99}, 33 (2012)
  [arXiv:1012.3983 [hep-th]].

\bibitem{Sieg:2010jt} 
  C.~Sieg,
  ``Review of AdS/CFT Integrability, Chapter I.2: The spectrum from perturbative gauge theory,''
  Lett.\ Math.\ Phys.\  {\bf 99}, 59 (2012)
  [arXiv:1012.3984 [hep-th]].


\bibitem{Siegel:1979wq} 
  W.~Siegel,
  ``Supersymmetric Dimensional Regularization via Dimensional Reduction,''
  Phys.\ Lett.\ B {\bf 84}, 193 (1979).

\bibitem{'tHooft:1982tz} 
  G.~'t Hooft,
  ``On the Convergence of Planar Diagram Expansions,''
  Commun.\ Math.\ Phys.\  {\bf 86}, 449 (1982).


\bibitem{Beisert:2003ys} 
  N.~Beisert,
  ``The $su(2|3)$ dynamic spin chain,''
  Nucl.\ Phys.\ B {\bf 682}, 487 (2004)
  [hep-th/0310252].


\bibitem{Beisert:2005tm} 
  N.~Beisert,
  ``The $SU(2|2)$ dynamic S-matrix,''
  Adv.\ Theor.\ Math.\ Phys.\  {\bf 12}, 945 (2008)
  [hep-th/0511082].
  
 
  
  
  
\bibitem{Kachru:1998ys} 
  S.~Kachru and E.~Silverstein,
  ``4-D conformal theories and strings on orbifolds,''
  Phys.\ Rev.\ Lett.\  {\bf 80}, 4855 (1998)
  [hep-th/9802183].

  
\bibitem{Lawrence:1998ja} 
  A.~E.~Lawrence, N.~Nekrasov and C.~Vafa,
  ``On conformal field theories in four-dimensions,''
  Nucl.\ Phys.\ B {\bf 533}, 199 (1998)
  [hep-th/9803015].

  
\bibitem{Klebanov:1999rd} 
  I.~R.~Klebanov and N.~A.~Nekrasov,
  ``Gravity duals of fractional branes and logarithmic RG flow,''
  Nucl.\ Phys.\ B {\bf 574}, 263 (2000)
  [hep-th/9911096].
  
  
\bibitem{Grisaru:1979wc} 
  M.~T.~Grisaru, W.~Siegel and M.~Rocek,
  ``Improved Methods for Supergraphs,''
  Nucl.\ Phys.\ B {\bf 159}, 429 (1979).

  
\bibitem{Grisaru:1982zh} 
  M.~T.~Grisaru and W.~Siegel,
  ``Supergraphity. 2. Manifestly Covariant Rules and Higher Loop Finiteness,''
  Nucl.\ Phys.\ B {\bf 201}, 292 (1982)
  [Erratum-ibid.\ B {\bf 206}, 496 (1982)].

  
\bibitem{Grisaru:1980nk} 
  M.~T.~Grisaru, M.~Rocek and W.~Siegel,
  ``Zero Three Loop beta Function in N=4 Superyang-Mills Theory,''
  Phys.\ Rev.\ Lett.\  {\bf 45}, 1063 (1980).


\bibitem{Stelle:1981gi} 
  K.~S.~Stelle,
  ``Extended Supercurrents And The Ultraviolet Finiteness Of N=4 Supersymmetric Yang-mills Theory,''
  In *London 1981, Proceedings, Quantum Structure Of Space and Time*, 337-361 and Paris Ec. Norm. Sup. - LPTENS 81-24 (81,REC.JAN.82) 26p

\bibitem{Howe:1982tm} 
  P.~S.~Howe, K.~S.~Stelle and P.~K.~Townsend,
  ``The Relaxed Hypermultiplet: An Unconstrained N=2 Superfield Theory,''
  Nucl.\ Phys.\ B {\bf 214}, 519 (1983).

\bibitem{Howe:1983wj} 
  P.~S.~Howe, K.~S.~Stelle and P.~C.~West,
  ``A Class of Finite Four-Dimensional Supersymmetric Field Theories,''
  Phys.\ Lett.\ B {\bf 124}, 55 (1983).


  
\bibitem{Howe:1983sr} 
  P.~S.~Howe, K.~S.~Stelle and P.~K.~Townsend,
  ``Miraculous Ultraviolet Cancellations in Supersymmetry Made Manifest,''
  Nucl.\ Phys.\ B {\bf 236}, 125 (1984).



  
\bibitem{Buchbinder:1997ya} 
  I.~L.~Buchbinder, E.~I.~Buchbinder, S.~M.~Kuzenko and B.~A.~Ovrut,
  ``The Background field method for N=2 superYang-Mills theories in harmonic superspace,''
  Phys.\ Lett.\ B {\bf 417}, 61 (1998)
  [hep-th/9704214].

  
\bibitem{Buchbinder:1998np} 
  I.~L.~Buchbinder and S.~M.~Kuzenko,
  ``Comments on the background field method in harmonic superspace: Nonholomorphic corrections in N=4 SYM,''
  Mod.\ Phys.\ Lett.\ A {\bf 13}, 1623 (1998)
  [hep-th/9804168].


\bibitem{Jain:2013hua} 
  D.~Jain and W.~Siegel,
  ``Improved Methods for Hypergraphs,''
  arXiv:1302.3277 [hep-th].


  
\bibitem{Gates:1983nr} 
  S.~J.~Gates, M.~T.~Grisaru, M.~Rocek and W.~Siegel,
  ``Superspace Or One Thousand and One Lessons in Supersymmetry,''
  Front.\ Phys.\  {\bf 58}, 1 (1983)
  [hep-th/0108200].


  
  
\bibitem{Buchbinder:1995uq} 
  I.~L.~Buchbinder and S.~M.~Kuzenko,
  ``Ideas and methods of supersymmetry and supergravity: A Walk through superspace,''
  Bristol, UK: IOP (1995) 640 p

\bibitem{Galperin:2001uw} 
  A.~S.~Galperin, E.~A.~Ivanov, V.~I.~Ogievetsky and E.~S.~Sokatchev,
  ``Harmonic superspace,''
  Cambridge, UK: Univ. Pr. (2001) 306 p
  


  
\bibitem{Grimm:1977xp} 
  R.~Grimm, M.~Sohnius and J.~Wess,
  ``Extended Supersymmetry and Gauge Theories,''
  Nucl.\ Phys.\ B {\bf 133}, 275 (1978).
  
\bibitem{Kovacs:1999rd} 
  S.~Kovacs,
  ``A Perturbative reanalysis of N=4 supersymmetric Yang-Mills theory,''
  Int.\ J.\ Mod.\ Phys.\ A {\bf 21}, 4555 (2006)
  [hep-th/9902047].

\bibitem{Abbott:1981ke} 
  L.~F.~Abbott,
  ``Introduction to the Background Field Method,''
  Acta Phys.\ Polon.\ B {\bf 13}, 33 (1982).
  
    

  
\bibitem{KlubergStern:1975hc} 
  H.~Kluberg-Stern and J.~B.~Zuber,
  ``Renormalization of Nonabelian Gauge Theories in a Background Field Gauge. 2. Gauge Invariant Operators,''
  Phys.\ Rev.\ D {\bf 12}, 3159 (1975).


\bibitem{Tarrach:1981bi} 
  R.~Tarrach,
  ``The Renormalization of FF,''
  Nucl.\ Phys.\ B {\bf 196}, 45 (1982).





%

\bibitem{deWit:1996kc} 
  B.~de Wit, M.~T.~Grisaru and M.~Rocek,
``Nonholomorphic corrections to the one-loop N=2 superYang-Mills action,''
  Phys.\ Lett.\ B {\bf 374}, 297 (1996)
  [hep-th/9601115].
  
  
\bibitem{Lindstrom:1989ne} 
  U.~Lindstrom and M.~Rocek,
  ``N=2 Superyang-mills Theory In Projective Superspace,''
  Commun.\ Math.\ Phys.\  {\bf 128}, 191 (1990).
  
\bibitem{Jain:2010gm} 
  D.~Jain and W.~Siegel,
  ``On Projective Hoops: Loops in Hyperspace,''
  Phys.\ Rev.\ D {\bf 83}, 105024 (2011)
  [arXiv:1012.3758 [hep-th]].
%


%
%
%




\bibitem{Lindstrom:1996xi} 
  U.~Lindstrom, F.~Gonzalez-Rey, M.~Rocek and R.~von Unge,
  ``On N=2 low-energy effective actions,''
  Phys.\ Lett.\ B {\bf 388}, 581 (1996)
  [hep-th/9607089].

\bibitem{Dine:1997nq} 
  M.~Dine and N.~Seiberg,
  ``Comments on higher derivative operators in some SUSY field theories,''
  Phys.\ Lett.\ B {\bf 409}, 239 (1997)
  [hep-th/9705057].


\bibitem{Kuzenko:2003wu} 
  S.~M.~Kuzenko and I.~N.~McArthur,
  ``On the two loop four derivative quantum corrections in 4-D N=2 superconformal field theories,''
  Nucl.\ Phys.\ B {\bf 683}, 3 (2004)
  [hep-th/0310025].
  
\bibitem{Banin:2002mf} 
  A.~T.~Banin, I.~L.~Buchbinder and N.~G.~Pletnev,
  ``On low-energy effective action in N=2 superYang-Mills theories on nonAbelian background,''
  Phys.\ Rev.\ D {\bf 66}, 045021 (2002)
  [hep-th/0205034].
  
  
\bibitem{Buchbinder:1998qd} 
  E.~I.~Buchbinder, I.~L.~Buchbinder and S.~M.~Kuzenko,
  ``Nonholomorphic effective potential in N=4 SU(n) SYM,''
  Phys.\ Lett.\ B {\bf 446}, 216 (1999)
  [hep-th/9810239].

  
  
\bibitem{Sieg:2005kd} 
  C.~Sieg and A.~Torrielli,
  ``Wrapping interactions and the genus expansion of the 2-point function of composite operators,''
  Nucl.\ Phys.\ B {\bf 723}, 3 (2005)
  [hep-th/0505071].
  


\bibitem{Smirnov:1999gc} 
  V.~A.~Smirnov,
  ``Analytical result for dimensionally regularized massless on shell double box,''
  Phys.\ Lett.\ B {\bf 460}, 397 (1999)
  [hep-ph/9905323].
  
\bibitem{Dolan:2002zh} 
  F.~A.~Dolan and H.~Osborn,
  ``On short and semi-short representations for four-dimensional superconformal symmetry,''
  Annals Phys.\  {\bf 307}, 41 (2003)
  [hep-th/0209056].

  
%
%
%
%
%
%


  
\bibitem{Andree:2010na} 
  R.~Andree and D.~Young,
  ``Wilson Loops in N=2 Superconformal Yang-Mills Theory,''
  JHEP {\bf 1009}, 095 (2010)
  [arXiv:1007.4923 [hep-th]].
  
\bibitem{Passerini:2011fe} 
  F.~Passerini and K.~Zarembo,
  ``Wilson Loops in N=2 Super-Yang-Mills from Matrix Model,''
  JHEP {\bf 1109}, 102 (2011)
  [Erratum-ibid.\  {\bf 1110}, 065 (2011)]
  [arXiv:1106.5763 [hep-th]].


\bibitem{Wenbin}
W.~Yan,  {Unpublished work}.

\bibitem{Mitev:2014yba} 
  V.~Mitev and E.~Pomoni,
  ``The Exact Effective Couplings of 4D N=2 gauge theories,''
  arXiv:1406.3629 [hep-th].



\bibitem{Belitsky:2003ys} 
  A.~V.~Belitsky, A.~S.~Gorsky and G.~P.~Korchemsky,
  ``Gauge / string duality for QCD conformal operators,''
  Nucl.\ Phys.\ B {\bf 667}, 3 (2003)
  [hep-th/0304028].

\bibitem{Brink:1982pd} 
  L.~Brink, O.~Lindgren and B.~E.~W.~Nilsson,
  ``N=4 Yang-Mills Theory on the Light Cone,''
  Nucl.\ Phys.\ B {\bf 212}, 401 (1983).


\bibitem{Mandelstam:1982cb} 
  S.~Mandelstam,
  ``Light Cone Superspace and the Ultraviolet Finiteness of the N=4 Model,''
  Nucl.\ Phys.\ B {\bf 213}, 149 (1983).
  
\bibitem{Ananth:2012tf} 
  S.~Ananth, S.~Kovacs and S.~Parikh,
  ``Gauge-invariant correlation functions in light-cone superspace,''
  JHEP {\bf 1205}, 096 (2012)
  [arXiv:1203.5376 [hep-th]].
  






  
  \end{thebibliography}

\end{document}